\documentstyle[prl,aps,preprint]{revtex}
\newcommand{\tr}{{\rm tr}}
\newcommand{\Tr}{{\rm Tr}}
\newcommand{\D}{{\rm D}}
\newcommand{\Dh}{{\cal D}}
\newcommand{\A}{{\cal A}}
\newcommand{\M}{{\cal M}}
\newcommand{\Mt}{\widetilde{\cal M}}
\newcommand{\Ct}{\widetilde{\rm C}}
\newcommand{\C}{{\rm C}}
\newcommand{\thru}[1]{\mathrel{\mathop{#1\!\!\!/}}}
\newcommand{\bfa}{\mbox{\boldmath $a$}}
\newcommand{\bfk}{\mbox{\boldmath $k$}} 
\newcommand{\bfp}{\mbox{\boldmath $p$}}
\newcommand{\bfx}{\mbox{\boldmath $x$}}
\newcommand{\bfA}{\mbox{\boldmath $A$}}
\newcommand{\bfD}{\mbox{\boldmath $D$}}
\newcommand{\bfgamma}{\mbox{\boldmath $\gamma$}}
\newcommand{\bftau}{\mbox{\boldmath $\tau$}}
\begin{document}

\draft
\tighten
\def\footnoterule{\kern-3pt \hrule width\hsize \kern3pt}

\title{Parity breaking in 2+1 dimensions and finite temperature}

\author{L.L. Salcedo}

\address{
{~} \\
Departamento de F\'{\i}sica Moderna \\
Universidad de Granada \\
E-18071 Granada, Spain
\ \\
\ \\
e-mail: salcedo@ugr.es \\
telephone: 34-958-246170 \\
FAX: 34-958-274208 
}

\maketitle

\thispagestyle{empty}

\begin{abstract}
An expansion in the number of spatial covariant derivatives is carried
out to compute the $\zeta$-function regularized effective action of
2+1-dimensional fermions at finite temperature in an arbitrary
non-Abelian background. The real and imaginary parts of the Euclidean
effective action are computed up to terms which are ultraviolet
finite. The expansion used preserves gauge and parity symmetries and
the correct multivaluation under large gauge transformations as well
as the correct parity anomaly are reproduced. The result is shown to
correctly reproduce known limiting cases, such as massless fermions,
zero temperature, and weak fields as well as exact results for some
Abelian configurations. Its connection with chiral symmetry is
discussed.
\end{abstract}

\vspace*{1cm}
PACS numbers:\ \ 11.10.Wx 11.10.Kk 11.15.-q 11.30.Er 11.30.RD

\vspace*{0.2cm}
Keywords:\ \ Finite temperature, Chern Simons, parity violation, three
dimensions, parity anomaly, gauge invariance, field theory, effective action

\vspace*{\fill}

\section{Introduction}
\label{sec:1}

Three-dimensional field theories find a direct physical application in
the context of condensed matter\cite{Lykken:1991vb} and also appear
as the high temperature limit of 3+1-dimensional theories
\cite{Ginsparg:1980ef}. From
the theoretical point of view, odd-dimensional field theories have
some remarkable properties. They are free from scale
anomalies \cite{Blau:1988kv} and so the corresponding
coupling constants do not run at one-loop
\cite{Hochberg:1998sv}. Gauge theories in odd-dimensional spaces
naturally admit a local term which is of topological nature, namely,
the Chern-Simons secondary characteristic class (CS). Such terms are
invariant under topologically small gauge transformations, induce a
parity breaking in the effective action and give a mass and spin to
the gauge boson\cite{Deser:1982vy}. Moreover, the so
called topological mass of the photon in QED$_3$ is not renormalized
beyond one loop
order\cite{Coleman:1985zi,Lykken:1991vb}. In the
non-Abelian case, the CS term may change by integer multiples of $2\pi
i$ under large gauge transformations, implying a quantization
condition on the mass to coupling constant ratio
\cite{Deser:1982vy}. The quantization of the CS
coefficient is in complete analogy with the quantization of the
topological Wess-Zumino-Witten action (WZW) in even-dimensional
space-times\cite{Wess:1971cm,Witten:1983tw}.

When fermions are coupled to the gauge fields, the CS term is induced
by quantum fluctuations
\cite{Redlich:1984kn,Niemi:1983rq,Alvarez-Gaume:1984ig}.
Preservation of gauge invariance requires then the presence of an
appropriate CS counter term which introduces a parity anomaly in the
effective action\cite{Redlich:1984kn}. The parity anomaly has the
same technical meaning as for axial anomalies in even dimensional
theories, that is, the anomalous term breaks a symmetry that was
present at the classical level. In the chiral case the fermions induce
a WZW term and enforcement of vector gauge invariance introduces an
axial anomaly
\cite{Adler:1969av}. The
dimensional ladder relating chiral, parity and gauge anomalies for
massless fermions in space-times of dimensions $D$+2, $D$+1 and $D$
has been elucidated in
\cite{Alvarez-Gaume:1985dr,Alvarez-Gaume:1985nf,Forte:1987em}. The
relation between axial anomalies in $D$+1 dimensions and effective
actions in $D$ dimensions using trace identities, both for zero and
finite temperature and density, has been studied in
\cite{Niemi:1983rq,Niemi:1985ux,Niemi:1986vz,Sissakian:1997cp}. It is
also noteworthy that induced CS terms appear in four dimensional
theories at finite temperature and density
\cite{Redlich:1985md}.

The exact parity-violating piece of the effective action for massless
fermions in odd-dimensional space-times was obtained in
\cite{Alvarez-Gaume:1985nf} in terms of the $\eta$-invariant.  This
result follows from the use of the $\zeta$-function regularization
\cite{Seeley:1967ea} which automatically preserves strict gauge
invariance under large and small gauge transformations. The
$\eta$-invariant measures the spectral asymmetry of the corresponding
Dirac operator, and through an application of the Atiyah-Patodi-Singer
index theorem \cite{Atiyah:1975,Alvarez-Gaume:1985nf}, it can be
related to the CS term corrected by the index of the Dirac operator
extended to one more dimension. Actually, the massless exact result
holds for zero as well as for finite temperature. Indeed, it applies
to fairly general space-time manifolds and, within the imaginary time
formulation, the finite temperature amounts to a S$^1$
compactification in the time direction.

For massive fermions there is no exact closed solution at zero
temperature but some exact results are known, including the presence
of a correctly quantized CS term (see
e.g.\cite{GamboaSaravi:1996aq,Salcedo:1996qy}). At finite temperature,
however, inconsistencies did appear in the first calculations of the
induced CS term in 2+1 dimensions both in the Abelian and non-Abelian
cases \cite{Niemi:1985ux,Ishikawa:1987zi}.  Since the CS term is a
local polynomial, the calculations to extract this term were usually
done using a combination of perturbative and derivative expansions
(see also \cite{Sissakian:1997cp} for a recent calculation of this
type at finite temperature and density, and \cite{Aitchison:1993md}
for perturbative calculations without the low momentum
approximation). The result of these calculations yields a CS term with
a coefficient which is a smooth function of the temperature and this
breaks invariance under large gauge transformations. The necessity of
gauge invariance was emphasized in\cite{Pisarski:1987gq}. In
\cite{Bralic:1996uj} it was argued that perhaps the exact (i.e., non
perturbative) coefficient is actually a stepwise function of the
temperature so that only properly quantized values appear and in any
case only those quantized values would contribute after the functional
integration on the gauge fields (this observation, however, does not
solve the problem when the gauge fields are external). Replacing the
CS by its gauge invariant version, namely, the $\eta$-invariant, does
not solve the problem either.  This is because this quantity is gauge
invariant at the price of displaying jumps of $\pm 2$ coming from its
index contribution.  However, if the spectrum of the Dirac operator is
discrete and sufficiently well behaved, the renormalized fermionic
determinant will be free from discontinuities under continuous
deformations of the gauge configuration, therefore the exact effective
action can only have jump discontinuities which are multiples of $2\pi
i$ (or multiples of $i\pi$ if some eigenvalue vanishes during the
deformation; this is the generic case for massless fermions). This
leads again to a quantization of the coefficient in the
$\eta$-invariant.

This apparent paradox has recently been solved: in
Ref.\cite{Dunne:1997yb} it is shown, in the exactly solvable
0+1-dimensional Abelian case, that the exact result is free of
pathologies and consistent with gauge invariance. Large gauge
transformations (which exists in this Abelian model due to the
compactification introduced by the temperature) mix different orders
of perturbation theory. In Refs. \cite{Deser:1997nv} the
2+1 dimensional case is considered. There it is shown that, using the
$\zeta$-function prescription, the effective action at finite
temperature can be regularized in a completely gauge invariant
manner. Furthermore, the imaginary part of Euclidean the effective
action is computed explicitly for some special configurations where
separation of the time and space variables is possible and the gauge
fields are effectively Abelian. In
Ref.\cite{Fosco:1997ei} another calculation is carried
out for such configurations, this time by integration of the chiral
anomaly of the associated two dimensional problem (the calculation is
extended to arbitrary odd dimensions in\cite{Fosco:1998cq}). These
results check invariance under both large and small gauge
transformations and display the expected parity anomaly. Related work,
both on odd and even dimensional space-times, prompted by the insights
in those papers can be found in
\cite{Deser:1997fu,Felipe:1997nx,Salcedo:1998tg}.

In the present work we study the Euclidean $\zeta$-function
regularized effective action of fermions in 2+1 dimensions in the
presence of arbitrary external non-Abelian gauge field configurations
and at finite temperature. The calculation is done not only for the
imaginary part, which contains the topological terms, but also for the
real part.

Besides the $\zeta$-function regularization, the main ingredients of
the calculation are, first, the use of a Wigner transformation to
represent the operators and second, an expansion in the number of
spatial covariant derivatives. The Wigner representation method was
introduced in\cite{Salcedo:1996qy} and is closely related to the
method of symbols for pseudo-differential operators. (An improved
symbols method and many references can be found
in\cite{Pletnev:1998yu}.) The method is adapted here to treat the
finite temperature case, and has been used also for 1+1- and
3+1-dimensional fermions at finite temperature in
\cite{Salcedo:1998tg} and for non-local Dirac operators in
\cite{RuizArriola:1998zi}. The combination of $\zeta$-function and
Wigner transformation yields a computational setup which is suitable
to obtain the effective action using different expansions.
\footnote{It should be noted, however, that the $\zeta$-function is
not particularly convenient to compute the effective action in
odd-dimensional space-times at zero temperature. This is because there
is no pathology free way to choose the branch cut in $z^s$ so as to
avoid both the spectrum of the Dirac operator, which is continuous
under most expansions, and the spectrum of the parity transformed
Dirac operator. This regularization is suitable at finite temperature
since then the spectrum becomes discrete for practical purposes.}

The other ingredient is the use of a derivative expansion. In order to
preserve gauge invariance, it is, of course, necessary to expand in
terms of the covariant derivatives. This would be sufficient at zero
temperature, however, at finite temperature because the frequencies
take on discrete values only, an expansion in the number of time-like
covariant derivatives breaks invariance under large gauge
transformations and so such derivatives have to be treated
non-perturbatively. Therefore, we will consider an expansion in the
number of spatial covariant derivatives only. The advantages of such
an expansion are, first, that it preserves gauge and parity
invariances, i.e., terms of different order do not mix under such
transformations, and second, each order increases the degree of
ultraviolet convergence. Therefore, computing the terms which are
ultraviolet divergent, as we will do, yields all contributions which
may contain anomalies and multivaluation under large gauge
transformations. In particular, topological terms are included there.
The remaining terms are strictly gauge and parity invariant.

For technical reasons, the calculation is carried out in the gauge
$\partial_0A_0=0$ which is shown to exist for any given configuration,
although it is not unique. We check that different $\partial_0A_0=0$
gauges yield compatible result. The calculation is not based on
integrating the axial anomaly in two dimensions or similar
ideas. Except for the use a particular gauge, it is a direct and
systematic calculation which uses only the definitions of
$\zeta$-function, derivative expansion, etc. This allows to obtain
both the topological and non-topological terms. The expansion can be
carried out to any order in principle, although, of course, higher
orders require an increasing amount of work which quickly becomes
prohibitive. As mentioned, in the present work we compute the terms
with ultraviolet divergences. This amounts to zero and two derivatives
for the real part and two derivatives for the imaginary part. We check
that our result is consistent with gauge invariance, more precisely,
it transforms as the CS term, and yields the correct parity anomaly,
which is shown to be temperature independent. Further, it reduces to
the result in \cite{Fosco:1997ei} for the same configurations, has
the correct zero temperature limit and reproduces the exact result for
massless fermions.

Another related expansion is also considered which consists in
carrying out a further expansion in powers of the time-like covariant
derivatives which are {\em inside} commutators (as will clear below,
the space-like covariant derivatives were already inside commutators
due to gauge invariance since the space is not compactified). This new
expansion still preserves gauge and parity invariances and higher
order terms are increasingly convergent. At leading order, the result
is a simpler expression which still contains all topological terms,
saturates the parity anomaly and possesses the same multivaluation of
the full effective action. This result also reproduces that
of\cite{Fosco:1997ei}, as well as the zero temperature and zero mass
limits. The same expansion has been used in\cite{Salcedo:1998tg} for
even-dimensional fermions at finite temperature.

The paper is organized as follows. In section~\ref{sec:2} some general
considerations are made regarding the effective action, its relevant
symmetries and the Wigner transformation method. In section
~\ref{sec:new.3} the 0+1-dimensional case is revised. This allows to
illustrate some exact results as well as the method before going to
the three-dimensional case. In section ~\ref{sec:summary} the main
results of the three-dimensional problem are presented. For the
leading term of the imaginary part, which is the finite temperature
version of the induced CS term, we check that it possesses all expected
correct properties under symmetry transformations and limiting cases.
In section \ref{sec:new.1} we present the explicit computation of the
effective action at lowest order, that is, without derivatives.  This
term is purely real.  Section \ref{sec:new.2} contains the computation
of the terms which are real and with two derivatives.  Section
\ref{sec:4} contains the similar calculation for the imaginary part.
Appendix~\ref{app:A} proves the existence of the $A_0$-stationary
gauge and finds the allowed gauge transformations within that gauge.
Finally, Appendix~\ref{app:B} introduces some mathematical results
used in the main text.

\section{General considerations}
\label{sec:2}

\subsection{The Dirac operator}
The most general Dirac operator in a flat three-dimensional Euclidean space
has the form
\begin{equation}
\D =\gamma_\mu D_\mu+M
\end{equation}
where $D_\mu=\partial_\mu+A_\mu$ is the covariant derivative and
$A_\mu(x)$ and $M(x)$ are matrices in some internal space to be
referred to as flavor. Our convention for the Dirac matrices will be
\begin{equation}
\gamma_\mu^\dagger=\gamma_\mu\,,\quad
\{\gamma_\mu,\gamma_\nu\}=2\delta_{\mu\nu}\,.
\end{equation}
In 2+1 dimensions there are two inequivalent irreducible
representations of the Dirac algebra labeled by $\eta=\pm 1$, defined
as $\gamma_0\gamma_1\gamma_2=i\eta$. For instance, $\gamma_0=\eta
\tau_3$, $\gamma_{1,2}=\tau_{1,2}$ where $\tau_i$ stand for the Pauli
matrices. To ensure unitarity in Minkowski space, $A_\mu$ is required
to be anti-Hermitian and $M$ Hermitian. At finite temperature $T$, the
Dirac operator $\D$ acts on the space of fermionic wave functions
$\psi(x)$, which are antiperiodic in the Euclidean time direction with
period $\beta=1/T$. Correspondingly, the external bosonic fields
$A_\mu$ and $M$ are periodic. Further restrictions on $\D$ comes by
imposing continuity in all involved functions (fermionic wave
functions, field configurations, gauge transformations, etc). In
addition, we will occasionally assume that the space is compactified
so that it is topologically a two dimensional sphere. This implies a
topology $M_3={\rm S}^2\times{\rm S}^1$ for the space-time. The class
of Dirac operators to be considered in each case must be a subset of
the above and should be sufficiently large as to be invariant under
the relevant symmetries of the problem.

\subsection{The effective action}
\label{subsec:2.b}
The unrenormalized partition functional is
\begin{equation}
Z(\D)= \int{\cal D}\psi{\cal D}\bar\psi\, e^{-\int d^3x\,\bar\psi\D\psi}\,.
\end{equation}
Formal integration of the fermions gives $Z={\rm Det}({\rm D})$ and thus
the unrenormalized effective action is
\begin{equation}
W_{\rm bare}(\D)= -\log Z(\D)= -\Tr\,\log(\D)\,.
\end{equation}
Upon regularization and renormalization of the ultraviolet
divergences, a well-defined and finite effective action $W$ is
obtained. However, the renormalization procedure is not unique and
thus there will be several functions $W(\D)$ all of them qualifying as
valid effective actions for the same original action $\int
d^3x\,\bar\psi\D\psi$. They are constrained to reproduce the same
ultraviolet finite terms since such terms can be computed without
regularization and are therefore unambiguous. This implies that the
fourth order variation should be common to all $W(\D)$. This is
because for arbitrary commuting variations of $\D$
\begin{equation}
\delta_1\delta_2\delta_3\delta_4 W(\D)=\Tr(\delta_1 \D\D^{-1}\delta_2
\D\D^{-1}\delta_3 \D\D^{-1}\delta_4 \D\D^{-1}) +\hbox{5 permutations}
\end{equation}
is ultraviolet finite in 2+1 dimensions. As a consequence, the
(renormalized) ultraviolet divergent terms, responsible of the
ambiguity in the renormalization, must vanish after a fourth order
variation. Thus the allowed ambiguity introduced by the
renormalization is a local polynomial action of canonical dimension at
most three constructed with the fields $M$, $A_\mu$ and their
derivatives. This is also clear in perturbation theory since all
Feynman diagrams become convergent after four derivatives in the
fields or the external momenta.\footnote{There is a subtlety at finite
temperature. In this case the frequency takes discrete values only and
this prevents to take a derivative with respect to it. Nevertheless,
since the ultraviolet convergence also increases by taking finite
differences, the argument applies also for finite temperature and the
allowed ambiguity is still a local polynomial.} Of course, in order to
compare the effective action for different Dirac operators, it is
essential to use the same renormalization for the whole class of Dirac
operators $\D$ considered.\footnote{As an extreme case of ignoring
this restriction, consider adding to the effective action a different
``constant'' for each $\D$. This would result in a completely
arbitrary function $W(\D)$.} Only in this case can be meaningful the
statement that $W(\D)$, as a function of $\D$, is unique modulo a
local polynomial.  All these remarks are well-known but they will be
needed later and non trivial consequences will be extracted from them.

A set of renormalizations are based on the spectrum of the Dirac
operator. Let $\D\psi_n=\lambda_n\psi_n$ be the eigenvalue equation
for $\D$. The ultraviolet divergent part of the spectrum comes from
the term $\gamma_\mu\partial_\mu$ of the Dirac operator and thus it
lies on $\pm i\infty$. In practice we will take $M(x)=m$, a constant
c-number, and in this case the spectrum will lie on the line $m+i{\rm
R}$. The bare effective action can be expressed as
\begin{equation}
W_{\rm bare}=-\sum_n\log(\lambda_n)\,.
\end{equation}
In this form, the sum over eigenvalues is ultraviolet divergent and
must be renormalized. However, the renormalization only affects the
ultraviolet part of the spectrum so any particular eigenvalue still
contributes with $-\log\lambda_n$ to the renormalized effective
action. This shows that qualitatively $W(\D)$ is to be understood as a
many-valuated function defined modulo $2\pi i$ on the manifold of
(sufficiently regular) Dirac operators to be denoted by $\M$. The
branching points correspond to the Dirac operators which are singular,
that is, with a zero eigenvalue.

A mathematically well founded choice for the renormalization
prescription, which we will adopt here, is the $\zeta$-function
version of $W$, namely,
\begin{equation}
W(\D)=-\sum_n\lambda_n^s\log\lambda_n\Big|_{s=0}\,.
\end{equation}
(The same branch is to be taken for the two functions $z^s$ and
$\log(z)$ and all the eigenvalues.)  This expression is to be understood as an analytical
extension on the variable $s$ from ${\rm Re}(s)<-3$ and a sufficient
condition to be well defined is that the eigenvalues are all on the
same Riemann sheet of the logarithm function and off the branch
cut\cite{Seeley:1967ea}.
Once the branch cut, $\Gamma$, for the function $-z^s\log(z)$ on the
complex plane is chosen, $W(\D)$ becomes completely well defined but
it will, of course, have a spurious discontinuity along the cut. More
precisely, the cut on the complex plane will produce on $\M$ a branch
cut stemming from each of the singular Dirac operators. The
corresponding cut manifold will be denoted by $\M_\Gamma$. On the
complex plane and upon analytical extension, two different choices of
the cut, $\Gamma_1$ and $\Gamma_2$, will yield the same function
$-z^s\log(z)$. This is no necessarily the case on $\M$. The two
choices of the cut will yield the same function $W(\D)$ upon
analytical extension if and only if $\Gamma_1$ and $\Gamma_2$ can be
deformed into each other crossing at most a finite number of
eigenvalues. Because the ultraviolet sector of the spectrum lies on
$\pm i\infty$ there are essentially only two inequivalent
prescriptions, namely, choosing the branch cut of the logarithm along
the negative real axis or along the positive real axis. The two
choices will be labeled by $\sigma=+1$ and $\sigma=-1$ and the two cut
manifolds by $\M_{\sigma=1}$ and $\M_{\sigma=-1}$, respectively.

Analogously to the theory of many-valued functions on the complex
plane, an alternative way to achieve both one-valuedness and
analyticity of $W(\D)$ is to work on the simply connected Riemann surface
manifold associated to $\M$ (i.e., its the universal covering
space) which will be denoted by $\Mt$. From the previous discussion it
follows that the two functions $W_{\sigma=\pm 1}(\D)$ defined on their
common Riemann surface $\Mt$ by analytical extension represent two
different ($\zeta$-function) renormalizations of the effective action
and thus must differ by a local polynomial function on $\Mt$.

Another comment refers to symmetries. As usual, a symmetry is defined
as any transformation of the Dirac operator which leaves invariant the
class of allowed Dirac operators and which can be compensated by a
corresponding transformation in the fermionic wave functions so that
the action $\int d^3x\,\bar\psi\D\psi$ remains unchanged. According to
this definition, symmetries leave invariant the classical effective
action (defined as the action evaluated at its minimum with respect to
the fermionic fields). A non-vanishing variation of the quantum
effective action implies a quantum anomaly for the symmetry. The
anomaly must be a local polynomial. Indeed, if $W(\D)$ is any
definition of the effective action and $\D^\Omega$ is the transformed
Dirac operator with $\Omega$ independent of $\D$, the function
$W(\D^\Omega)$ also qualifies as an admissible definition of the
effective action of $\D$. Therefore, the anomaly $W(\D^\Omega)-W(\D)$
must be a local polynomial. Instances of this are the chiral anomaly
in even dimensional space-times or the parity anomaly in odd
dimensions. It also applies to the $2\pi in$ multivaluation under
large gauge transformations.\footnote{It follows that it is always
possible to define effective actions free from anomalies associated to
compact groups, namely, by taking an average over all symmetry
transformed configurations. This applies in particular to parity
transformations, however, the average of the two effective actions
defined modulo $2\pi i$ will have an ambiguity $i\pi$ in general, so
gauge invariance is not ensured.}

In practice, the $\zeta$-function renormalization is carried out by
introducing the quantity
\begin{equation}
\Omega_s(\D)=\Tr(\D^s) = \sum_n\lambda_n^s\,,
\end{equation}
which is ultraviolet finite if Re$(s)<-3$ and a meromorphic function
of $s$ with simple poles in $s=-3,-2,-1$
\cite{Seeley:1967ea}. Then, the
$\zeta$-function prescription is
\begin{equation}
W(\D)= -\frac{d}{ds}\Omega_s(\D)\Big|_{s=0}\,.
\end{equation}
The precise definition of the function $z^s$ requires cutting the
complex plane along a ray characterized by an angle $\theta$. Applying
Cauchy's theorem\cite{Seeley:1967ea}
\begin{equation}
\Omega_{s,\Gamma}(\D)= -\Tr\int_\Gamma\frac{dz}{2\pi
i}\frac{z^s}{\D-z}\,.
\label{eq:new.10}
\end{equation}
The trace $\Tr$ is to be taken in the fermionic Hilbert space which is the
tensor product of space, time, flavor and Dirac. The integration path
$\Gamma$ follows the ray $\theta$ starting from infinity, encircles
zero clockwise and goes back to infinity along the ray $\theta-2\pi$,
so that if the spectrum of the operator $\D$ were bounded $\Gamma$
could be deformed to enclose it anticlockwise. The choices
$\theta=\pi$ and $2\pi$ correspond to $\sigma=+1$ and $-1$
respectively. As noted, any value $\pi/2 < \theta < 3\pi/2$ gives the
same function $\Omega_{s,\sigma=+1}$ on the Riemann, surface, and
similarly any $3\pi/2< \theta< 5\pi/2$ yields the same function
$\Omega_{s,\sigma=-1}$. Besides, The same effective action follows
from taking $\theta$ or $\theta+2\pi n$ for integer $n$. (See
Appendix~\ref{app:B}.) 

\subsection{Gauge transformations}
\label{subsec:2.c}
Gauge transformations are defined as $\D^U=U^{-1}\D U$, where $U(x)$
is a matrix valued function acting as a multiplicative operator on the
fermionic wave functions. More explicitly,
$\D^U=\gamma_\mu(\partial_\mu+A^U_\mu)+M^U$, with
\begin{equation}
A^U_\mu(x)= U^{-1}(x)(\partial_\mu+A_\mu(x))U(x)\,,\quad
M^U(x)=U^{-1}(x)M(x)U(x)\,.
\end{equation}
$U(x)$ belongs to some gauge group $G$ which is a subgroup of
U$(N_f)$, $N_f$ being the number of flavors. Correspondingly $A_\mu$
must be in the Lie algebra of $G$ and likewise the class of matrices
$M(x)$ must also be closed under gauge transformations. (This is
trivially satisfied if $M$ is just a c-number.) Besides, $U(x)$ must
be continuous on the space-time manifold and in particular periodic as
a function of $x_0$.

Being similarity transformations, gauge transformations leave the
spectrum of $\D$ invariant. As a consequence, the functions
$W_{\sigma=\pm 1}(\D)$ defined on the cut manifolds $\M_{\sigma=\pm
1}$ are strictly gauge invariant by construction since these functions
depend solely on the spectrum of $\D$. In this sense, the
$\zeta$-function renormalization prescription trivially preserves
gauge invariance. On the Riemann surface $\Mt$ the statement is less
trivial. Let $\D$ be continuously deformed along a path connecting two
field configurations which are related by a gauge transformation. This
will induce a corresponding bounded path on the complex plane for each
eigenvalue and, since the spectrum is unchanged at the end, the net
effect on the eigenvalues can be at most a permutation. During its
walk, a finite number of eigenvalues will cross the branch cut, each
time adding a net $\pm 2\pi i$, to the effective
action.\footnote{Recall that each particular eigenvalue adds
$-\log\lambda_n$ to the {\em renormalized} effective action; the
renormalization, $\lim_{s\to 0}\sum_n\lambda_n^s\log\lambda_n$, leaves
no trace on any given bounded region of the complex plane.} In summary,
the functions $W_{\sigma=\pm 1}(\D)$ extended to $\Mt$ are gauge
invariant modulo $2\pi i$. It should be noted that, by continuity, the
effect of a gauge transformation may only depend on the homotopy class
of the gauge transformation and on the homotopy class of the initial
field configuration. In particular, only topologically large gauge
transformations may change the effective action by a non vanishing
integer multiple of $2\pi i$. We are assuming that no branching point
lies on the path followed by the eigenvalues. In the massless case
this cannot be avoided since the spectrum is purely imaginary. In this
particular case the variation in $W_{\sigma=\pm 1}(\D)$ is a multiple
of $i\pi$ instead.

For convenience, the calculation of the effective action will be
carried out in an $A_0$-stationary gauge, i.e., a gauge such that
$\partial_0 A_0(x)=0$. Such a gauge always exits for any given gauge
configuration, therefore there are no loss of generality by making
this choice. This result is proven in Appendix~\ref{app:A}. In the
same appendix, it is shown that the gauge transformation needed to
bring a gauge field configuration to an $A_0$-stationary gauge can
always be chosen to be topologically small in the Abelian case. On the
other hand, for simply connected gauge groups large transformations
may be necessary, depending on the given gauge configuration.

To check gauge invariance of our calculation, it will also be
necessary to find the most general gauge transformations which
preserve the gauge fixing condition. As shown in Appendix~\ref{app:A},
these transformations have the form
\begin{equation}
U(x)= \exp(-x_0A_0(\bfx))\exp(x_0(A_0(\bfx)+\Lambda(\bfx)))U_0(\bfx)\,.
\label{eq:A1}
\end{equation}
Here $U_0(\bfx)$ is an arbitrary time-independent gauge
transformation. $\Lambda$ is a time-independent function taking values
in the Lie algebra of the gauge group. It is only restricted by
continuity of $U(x)$, which requires
\begin{equation}
\exp(\beta A_0(\bfx))=\exp(\beta (A_0(\bfx)+\Lambda(\bfx)))\,.
\label{eq:A2}
\end{equation}
Therefore, the most general gauge transformation preserving the
$A_0$-stationarity condition is composed of a particular
time-dependent transformation subject to the condition in
eq.~(\ref{eq:A2}) followed by a stationary gauge transformation. The
corresponding zeroth component of the transformed field takes the form
\begin{equation}
A^U_0(\bfx)=U_0^{-1}(\bfx)(A_0(\bfx)+\Lambda(\bfx))U_0(\bfx)\,.
\label{eq:3}
\end{equation}
The time-dependent transformations $e^{-x_0A_0}e^{x_0(A_0+\Lambda)}$
will be called discrete gauge transformations for the stationary field
$A_0$ (since generically they form a discrete set due to
eq.~(\ref{eq:A2})).

In general the condition on $\Lambda$ can be simplified. From
eq.~(\ref{eq:A2}), it follows that both $A_0(\bfx)$ and $\Lambda(\bfx)$
must commute with $\exp(\beta A_0(\bfx))$. If the spectrum of
$\exp(\beta A_0(\bfx))$ is nowhere degenerated, this unitary matrix
takes a diagonal form in an $\bfx$-dependent basis in flavor space
that is essentially unique, thus $A_0(\bfx)$ and $\Lambda(\bfx)$ must
also be diagonal in the same basis and therefore they commute with
each other. In this case the condition on $\Lambda$ becomes
\begin{equation}
[A_0(\bfx),\Lambda(\bfx)]=0\,,\quad \exp(\beta \Lambda(\bfx))=1\,.
\label{eq:A3}
\end{equation}
The second condition means that the eigenvalues of $\Lambda(\bfx)$ are
of the form $\lambda_j=2\pi in_j/\beta$, for integer $n_j$. Such
integers are ${\bfx}$-independent by continuity. Correspondingly, in
this case, the allowed discrete transformations take the simpler form
\begin{equation}
U(x)= \exp(x_0\Lambda(\bfx))\,.
\label{eq:9}
\end{equation}
This form is the generic one. It applies if the spectrum of
$\exp(\beta A_0(\bfx))$ is nowhere degenerated, or if the points of
degeneracy can be resolved by continuity.

The stationary gauge transformations are always topologically small
since the second homotopy group is trivial for any Lie group. Whether
the discrete transformations are topologically large or not depends on
several factors. If the gauge group is not simply connected, for
instance the Abelian case, U(1), and $\Lambda$ is not zero, such
transformations are necessarily large, since the loop around $S^1$ for
each $\bfx$ cannot be contracted to the identity. If the gauge group
is simply connected, for instance SU($N_f$), the discrete
transformation may be large or small depending on $\Lambda(\bfx)$. For
instance, if $\Lambda$ is $\bfx$-independent, or becomes so after a
continuous deformation, the discrete transformation will be
topologically small. In particular, this holds if $\Lambda(\bfx)$ can
be diagonalized using a similarity transformation which is continuous
on S$^2$, since a diagonal $\Lambda$ must be $\bfx$-independent due to
eq.~(\ref{eq:A3}). On the other hand, to see that there are large
discrete transformation in general, consider the discrete
transformations from S$^2\times$S$^1$ into SU(2),
\begin{equation}
U(\bfx,x_0)= \exp\left(\frac{2\pi nx_0}{\beta}
i\bftau\bfx\right)\,,\quad n\in {\rm Z}
\end{equation}
where $\bftau$ are the Pauli matrices and $\bfx$ belongs to the unit
sphere S$^2$ in R$^3$. $U(x)$ covers SU(2) $2n$ times (it is $2n$ to
one and it maps a positively oriented volume element on
S$^2\times$S$^1$ into a positively oriented volume element on
SU(2)). Equivalently, one can compute the normalized
Wess-Zumino-Witten 3-form (relevant in chiral theories in 1+1
dimensions) which is closed and thus invariant under deformations,
\begin{equation}
\Gamma_{\rm WZW}(U)= -\frac{i}{12\pi}\int_{M_3}\tr(R^3)=-4i\pi n\,, \quad
R=U^{-1}dU.
\end{equation}
Therefore, such discrete transformations are all homotopically
inequivalent and large for non vanishing $n$. The Wess-Zumino-Witten
action appears naturally in this context since the Chern-Simons action
\begin{equation}
W_{\rm CS}(A)= \frac{i}{4\pi}\int_{M_3}\tr(AF-\frac{1}{3}A^3)\,,
\end{equation}
(where $F=dA+A^2$) evaluated for a pure gauge field $A=U^{-1}dU=R$ is
just $\Gamma_{\rm WZW}(U)$. As noted in the introduction, the
Chern-Simons action can only change under gauge transformations which
are topologically large.

\subsection{Parity transformation}
\label{subsec:2.d}
The parity transformed of $\D$ will be defined as
$\D^P=\gamma_\mu(\partial_\mu+A^P_\mu)+M^P$ with
\begin{equation}
A_0^P(x)=-A_0(x^P)\,,\quad \bfA(x)=+\bfA(x^P)\,,\quad
M^P(x)=-M(x^P)\,, \quad x^P=(-x_0,\bfx)\,.
\end{equation}
This is the definition taken in\cite{Alvarez-Gaume:1985nf}. All
definitions of parity consist in reversing an odd number of
coordinates and are related by a similarity transformation thereby
being equivalent.

The fermionic field transforms as $\psi^P(x)=\gamma_0\psi(x^P)$ and
the eigenvalue equation becomes $\D^P\psi^P_n=-\lambda_n\psi^P_n$.
Since the eigenvalues are changed, $W(\D)$ needs not be parity invariant
within the $\zeta$-function regularization. This allows for a parity
anomaly which is defined as 
\begin{equation}
\A_P(\D)= W(\D)-W(\D^P)\,.
\end{equation}
As noted at the end of subsection~\ref{subsec:2.b}, the parity anomaly
is due to the ultraviolet divergences present at the quantum level and
hence is a local polynomial in the fields. The same result follows
from noting that the two choices for the cut, $\sigma=\pm 1$, are
related by a parity transformation and thus
\begin{equation}
W_\sigma(\D^P)= W_{-\sigma}(\D)\,.
\label{eq:18b}
\end{equation}
As a consequence, the anomaly can also be written as $\A_P(\D)=
W_\sigma(\D)-W_{-\sigma}(\D)$, which, as noted above, is a local
polynomial action on $\Mt$.

Under parity, the action can be split into a parity preserving part
plus a parity violating part, or equivalently, a $\sigma$-even plus a
$\sigma$-odd component
\begin{equation}
W_{\rm even}(\D)=\frac{1}{2}(W_\sigma(\D)+W_{-\sigma}(\D))
\,,\quad
W_{\rm odd}(\D)=\frac{1}{2}\A_P(\D)\,.
\end{equation}
Since $W_{\rm odd}$ is just a local polynomial, it can be removed by
counterterms to end up with a parity preserving effective
action. However, this can be incompatible with gauge invariance. By
construction $W_\sigma$ and $\A_P$ are gauge invariant modulo $2\pi
i$, but $W_{\rm even}$ and $W_{\rm odd}$ can separately change by a
multiple of $i\pi$ in general. When this happens, adding counterterms
to remove $W_{\rm odd}$ spoils gauge invariance of the effective
action. Parity and gauge invariances cannot be both enforced in
general\cite{Redlich:1984kn}.

Because the parity anomaly is due to divergences in the ultraviolet
sector of the theory, it is to be expected that it is essentially
unchanged by introducing a finite temperature, which only affects the
infrared sector through the periodic boundary conditions. The same
argument applies for other quantum anomalies. In particular, it is
well established in the literature (e.g. \cite{GomezNicola:1994vq})
that the axial anomaly takes the Bardeen form independently of the
temperature. Remarkably, the analogous result can be proven for the
parity anomaly, even without a detailed calculation, namely, the form
of the essential parity anomaly at finite temperature is given by the
Chern-Simons action. By essential it is meant the part of the anomaly
which is not removable by gauge invariant counter terms (i.e., the
analogous of the Bardeen anomaly in the chiral case). To prove this
result, we start with the parity anomaly at zero temperature. It is
given by the Chern-Simons action which, of course, is a local
polynomial, parity odd and gauge invariant modulo $2\pi i$. Therefore,
at zero temperature, $W_{\rm odd}=\pm\frac{1}{2}W_{\rm CS}$. At finite
temperature, $W_{\rm odd}$ will be given by $\pm\frac{1}{2}W_{\rm CS}$
plus a local polynomial with a smooth temperature dependence. By
continuity this local polynomial must be strictly gauge invariant (not
just modulo $i\pi$) and thus it can be removed by a counterterm
without destroying gauge invariance, leaving $\A_P=\pm W_{\rm CS}$ at any
temperature. This conclusion is confirmed by our detailed calculation
below and also follows from eq.~(5) of \cite{Deser:1997nv} (noting the
different definitions of parity violating terms in both works). The
same argument applies to any other smooth dependence such as that
introduced by a finite density or a change in the mass of the fermion.

\subsection{Pseudo-parity transformation}
It is also of interest to introduce a pseudo-parity
transformation\cite{Witten:1983tw}, defined as
\begin{equation}
A_0(x)\to -A_0(x^P)\,,\quad
{\bfA}(x)\to +\bfA(x^P)\,,\quad
M(x)\to +M(x^P)\,.
\end{equation}
It is clear that the pseudo-parity odd component of the effective
action is that with an odd number of zeroth Lorentz indices and thus
is the component containing the Levi-Civita pseudo-tensor. It also
corresponds to the imaginary part of the Euclidean effective
action. The CS action is pseudo-parity odd. The $\zeta$-function
prescription yields a parity anomaly both in the pseudo-parity even
and odd components, but only the latter is essential in the sense that
it cannot be removed without breaking gauge invariance. This is
because the even component can be renormalized preserving gauge and
parity invariances
simultaneously\cite{Alvarez-Gaume:1985dr,Alvarez-Gaume:1985nf}.
Note that in the massless case, parity and pseudo-parity
transformations coincide.

\subsection{Wigner transformation method}
The (asymmetric) Wigner transformation method is described in great
detail in\cite{Salcedo:1996qy}. The corresponding formula for the
zero-temperature (i.e., uncompactified) case is given in that reference
and only a variation is needed to cope with the compactification in
the zeroth direction. We concentrate on the Hilbert spaces of periodic
and antiperiodic wave functions of time defined on S$^1$. Let us
introduce the fermionic and bosonic Matsubara frequencies
\begin{equation}
\omega_n= \frac{2\pi i}{\beta}(n+\frac{1}{2})\,,\quad
\kappa_n= \frac{2\pi in}{\beta} \,.
\end{equation}
The states $|\omega\rangle= e^{\omega x_0}$ are periodic if
$\omega=\kappa_n$ and antiperiodic if $\omega=\omega_n$. In both cases
they are normalized to $\sqrt{\beta}$. A general operator acting on
the antiperiodic space will be of the form
$f(\hat{x}_0,\hat{\partial}_0)$ where $\hat{x}_0$ and
$\hat{\partial}_0$ denote the time and frequency operators
respectively, and $f$ is a periodic function of $\hat{x}_0$.  Then the
trace of $f$ is given by
\begin{eqnarray}
\Tr(f(\hat{x}_0,\hat{\partial}_0)) &=&
\frac{1}{\beta}\sum_{n\in{\rm Z}}\langle\omega_n|
f(\hat{x}_0,\hat{\partial}_0)|\omega_n\rangle
\nonumber\\
&=& \frac{1}{\beta}\sum_{n\in{\rm Z}}\langle 0|e^{-\omega_n \hat{x}_0}
f(\hat{x}_0,\hat{\partial}_0)
e^{\omega_n \hat{x}_0}|0\rangle \nonumber\\
&=& \frac{1}{\beta}\sum_{n\in{\rm Z}}\langle 0|
f(\hat{x}_0,\hat{\partial}_0 + \omega_n) |0\rangle \,.
\end{eqnarray}
Here $|0\rangle$ is the zero frequency state.  Note that $|0\rangle$
is periodic rather than antiperiodic.  For the non compact coordinates
the treatment is entirely similar, $D_i$ is replaced by $D_i+p_i$,
there is an integral over momenta and $|0\rangle$ is the zero momentum
state\cite{Salcedo:1996qy}. Therefore, using eq.~(\ref{eq:new.10}),
the $\zeta$-function can then be written as
\begin{equation}
\Omega_{s,\Gamma}(\D)= -\int\frac{d^2k}{(2\pi)^2} \frac{1}{\beta}\sum_n
\int_\Gamma\frac{dz}{2\pi i}z^s\tr \langle
0|\frac{1}{\thru{p}+\D-z}|0\rangle.
\label{eq:7}
\end{equation}
Here $\bfp=i\bfk$ (where $\bfk\in{\rm R}^2$), $p_0=\omega_n$ and
$|0\rangle$ represents the spinless and flavorless zero
energy-momentum state normalized as $\langle x|0\rangle=1$. In
particular $\partial_\mu|0\rangle= \langle 0|\partial_\mu= 0$, and
also $\langle 0|0\rangle=\int d^3x$, and moreover, if $f(\hat{x})$ is
a multiplicative operator, i.e., it does not contain derivatives,
\begin{equation}
\langle 0|f(\hat{x})|0\rangle=\int d^3x f(x)\,.
\end{equation}

The formula just given for $\Omega_{s,\Gamma}(\D)$ is completely
finite in the ultraviolet sector and exact if a space-time topology
R$^2\times$S$^1$ is assumed.

\subsection{Covariant derivative expansion}
\label{subsec:2.g}

We will find convenient to use an $A_0$-stationary gauge to compute
the effective action. This is allowed due to gauge invariance,
however, since the $A_0$-stationary gauge is not unique, we will have
to check that the result is the same for any of the gauge copies,
i.e., invariant under time-independent and discrete gauge
transformations. It should be noted that the integrand in
eq.~(\ref{eq:7}) is not directly gauge invariant due to the presence
of $|0\rangle$; gauge invariance requires integration over momentum
and sum over frequencies\cite{Salcedo:1996qy}. Roughly speaking,
invariance under time independent transformations means that $D_i$
appears only inside commutators and this is equivalent to say that the
result is unchanged if $D_i$ is replaced by $D_i+a_i$, $\bfa$ being an
arbitrary $\bfx$-independent c-number, but precisely this invariance
is ensured by the integration over momentum since $\bfa$ can always be
absorbed in $\bfp$. Similarly, invariance under the discrete gauge
transformations, $A_0\to A_0+\Lambda$, is ensured by the sum over
discrete frequencies, since $\Lambda$ can be absorbed in $\omega_n$
(recall that the spectrum of $\Lambda$ is quantized). We note that a
recently proposed technique, due to Pletnev and
Banin\cite{Pletnev:1998yu}, allows to bring the expression to one
where $D_i$ appears inside commutators from the beginning, i.e., prior
to momentum integration.  Presently this technique applies to
covariant derivatives in non compact directions.

In order to carry out the traces, sums and integrations implied in
eq.~(\ref{eq:7}) we will perform a formal expansion in powers of the
spatial covariant derivatives, $D_i$, $i=1,2$.  However, it is
important to remark that such an expansion is not only formal; it is
meaningful for the effective action functional itself beyond the
particular computational procedure followed. It actually means to
change the bosonic field configuration to $A_0(\lambda\bfx,x_0)$,
$\lambda\bfA(\lambda\bfx,x_0)$ and $M(\lambda\bfx,x_0)$, and then
count powers of $\lambda$. Since this scaling preserves full gauge
invariance (namely, by scaling $U$ too) the gradient expansion is
gauge invariant order by order. Formally, the scaling corresponds to
the replacement $D_i\to \lambda D_i$ and gauge invariance is preserved
since $\lambda a_i$ can still be compensated by a change of variables
in $k_i$. The situation is completely different for a formal expansion
in powers of $D_0$. First of all, it would only be formal, since
scaling $x_0\to \lambda x_0$ violates the boundary conditions for most
field configurations (because $\beta$ is not scaled). At a formal
level, scaling $D_0$ would imply a similar scaling in $\Lambda$ which
will no longer be properly quantized and it could not be compensated
by a shift in $\omega_n$.

Due to rotational invariance, the proposed gradient expansion contains
even orders only. In this work we explicitly work out the zeroth and
second order terms, $W_0(\D)$ and $W_2(\D)$ respectively. Since all
higher order terms are ultraviolet finite and the different orders do
not mix under gauge or parity transformations, all anomalies are
contained in these two terms. The pseudo-parity odd component starts
at second order since the Levi-Civita pseudo-tensor requires the
presence of at least two gradients. Thus the leading pseudo-parity odd
term, $W^-_2(\D)$, is responsible for all essential anomalies in the
effective action, whereas $W_0$ and $W_2^+$ may contain at most
removable anomalies.

There is another expansion which is also of great interest in the
present context. As will be clear from the calculation, the final
result at each order in the spatial gradient expansion comes as a
function of $A_0$ and $\Dh_0$ which is defined as $[D_0,\,]$. (Note
that $A_0$ and $\Dh_0$ commute in a $A_0$-stationary gauge.) It is the
expansion in powers of $A_0$ the one that would break gauge invariance
under discrete gauge transformations, and so $A_0$ should be treated
non perturbatively. On the other hand, expanding in powers of $\Dh_0$
does not break any symmetry of the problem (excepting, of course,
Lorentz invariance in the limit of zero temperature). Each extra power
of $\Dh_0$ or $D_i$ increases the degree of convergence in the
ultraviolet sector. Therefore, at finite temperature, it is natural to
consider a double expansion in powers of $D_i$ and $\Dh_0$ which play
a similar role as a standard space-time gradient expansion which is
often used at zero temperature. (In fact, $D_i$ always appears inside
of commutators and so the expansion in powers of $D_i$ can be seen
also as an expansion in powers of $\Dh_i=[D_i,\,]$). This same idea
has been applied in the case of 3+1-dimensional fermions at finite
temperature \cite{Salcedo:1998tg}.

\section{The 0+1-dimensional case}
\label{sec:new.3}

Before going to 2+1 dimensions, we will study the simpler
0+1-dimensional case. This will allow us to illustrate some of the
previous remarks as well as the method.

In 0+1 dimensions, the Dirac operator is $\gamma_0D_0+m$ with
$\gamma_0=\eta=\pm 1$, a $1\times 1$ dimensional matrix in Dirac
space. We have already taken $M(x)=m$, a constant c-number. We start
by fixing the gauge to be $A_0$-stationary. For convenience we will
use units $\beta=2$. A direct application of the Wigner transformation
formula (i.e., eq.~(\ref{eq:7}) adapted to 0+1 dimensions) gives:
\begin{eqnarray}
\Omega_{s,\Gamma}(\D) &=& -\frac{1}{2}\sum_n\int_\Gamma
\frac{dz}{2\pi i}z^s\tr\langle 0|\frac{1}{\gamma_0 (\omega_n+D_0)+m-z}
|0\rangle
\nonumber \\
&=& -\sum_n\int_\Gamma
\frac{dz}{2\pi i}z^s\tr\left(\frac{1}{\eta (\omega_n+A_0)+m-z}\right) \,.
\end{eqnarray}
We have replaced $D_0$ by $A_0$ since everything is time independent
and $\partial_0$ vanishes on $|0\rangle$, and $\langle
0|0\rangle=\beta=2$ has been used.

In order to sum over frequencies, we first transform the series in
$\Omega_{s,\Gamma}(\D)$ into a convergent one by performing a subtraction
\begin{equation}
\Omega_{s,\Gamma}(\D) = -\int_\Gamma
\frac{dz}{2\pi i}z^s\tr\sum_n\left(
\frac{1}{\eta (\omega_n+A_0)+m-z}
-\frac{1}{\eta (\omega_n+A_0)+m}
\right) \,.
\label{eq:new.30}
\end{equation}
This is justified due to the identity
\begin{equation}
\int_\Gamma\frac{dz}{2\pi i}z^sq(z) = 0 \,,
\label{eq:42}
\end{equation}
for any polynomial $q(z)$, which comes from closing $\Gamma$ with the
circumference at infinity. This and similar formulas are to be
understood upon analytical extension from sufficiently negative ${\rm
Re}(s)$. Then, the following formula can be used
\begin{equation}
\sum_n\left(\frac{1}{x_1+\omega_n}-\frac{1}{x_2+\omega_n}\right)
=\tanh(x_1)-\tanh(x_2)\,
\label{eq:new.1}
\end{equation}
and the sum over frequencies yields
\begin{equation}
\Omega_{s,\Gamma}(\D)= -\int_\Gamma
\frac{dz}{2\pi i}z^s\,\tr\tanh(m+\eta A_0-z)\,.
\label{eq:new.2}
\end{equation}

The final step is to carry out the integration over $z$ and apply
$-d/ds|_{s=0}$ to obtain the effective action. This can be done using
the functions
\begin{equation}
\Omega_\Gamma(\omega,s) =
 -\int_\Gamma
\frac{dz}{2\pi i}z^s
\tanh(\omega-z) \,.
\label{eq:new.7}
\end{equation}
The properties of these and related functions are summarized in
Appendix~\ref{app:B}. As shown there
\begin{eqnarray}
\Omega_\sigma(\omega,0) &=& 0\nonumber \\
\Omega^\prime_\sigma(\omega,0) &=& \log(e^{-2\sigma\omega}+1)\,.
\end{eqnarray}
Here the prime refers to derivative with respect to $s$ and
$\sigma=\pm 1$ refers to the two possible inequivalent choices of
$\Gamma$. Two remarks can be made regarding the first of these
equations. First, the vanishing of $\Omega_\sigma(\omega,0)$
guarantees that taking $\Gamma$ along the ray $\theta$ or $\theta+2\pi
n$ yields the same final result for the effective action. Second, it
also implies that there is no scale anomaly in 0+1
dimensions. Indeed, the scale anomaly can be associated to a non
trivial dependence of the effective action on some mass parameter
$M_0$ introduced by using $(z/M_0)^s$ instead of $z^s$ in the
definition of the $\zeta$-function. Such dependence cancels in odd
dimensions\cite{Blau:1988kv}.

The final exact result for the effective action in 0+1 dimensions,
restoring arbitrary units, is thus
\begin{equation}
W_\sigma(\D)= -\tr\log(e^{-\sigma\beta(m+\eta A_0)}+1)\,.
\end{equation}
Recalling that under gauge transformations the eigenvalues of $A_0$
may change at most by integer multiples of $2\pi i/\beta$ it follows
that $W_\sigma(\D)$ is manifestly gauge invariant modulo $2\pi i$ on
the Riemann surface $\Mt$ and strictly gauge invariant on each Riemann
sheet $\M_\sigma$. This latter property also follows from
eq.~(\ref{eq:new.2}) since the integrand there is strictly periodic;
this is a direct consequence of the sum over $\omega_n$. As noted
in\cite{Dunne:1997yb}, a perturbative expansion in powers of $A_0$
would display a spurious breaking of gauge invariance at finite
temperature. This is because the perturbative condition, $A_0$ small,
is not preserved by large gauge transformations. This simply means
that a truncated series expansion of a periodic function is not itself
a periodic function in general.

The behavior under parity is better exposed by introducing the
functions $\phi_n$ of Appendix~\ref{app:B}. In the present case
\begin{eqnarray}
\Omega^\prime_\sigma(\omega,0) &=& \phi_0(\omega)-\sigma\omega\,, \nonumber \\
\phi_0(\omega)&=& \log(2\cosh(\omega)) \,.
\end{eqnarray}
The parity transformation corresponds here to $\omega\to -\omega$. In
agreement with our previous considerations, the term that breaks
parity, $-\sigma\omega$, is a polynomial. Also, a parity
transformation is compensated by a change $\sigma\to -\sigma$, in
agreement with eq.~(\ref{eq:18b}). The effective action can then be
rewritten as
\begin{equation}
W_\sigma(\D)= -\tr\log(2\cosh(\frac{\beta}{2}(m+\eta A_0)))+
\eta\sigma\frac{\beta}{2}\tr(A_0) +\sigma N_f \frac{\beta}{2}m\,.
\end{equation}
The first term coincides with the result in\cite{Dunne:1997yb} after
subtraction of a $A_0$-independent term (namely, the same expression
with $A_0=0$). It preserves parity but may break gauge invariance
since it changes by integer multiples of $i\pi$. This is compensated
by the second term which is odd under pseudo-parity and breaks both
gauge and parity symmetries. It is a local polynomial that introduces
a parity anomaly. It cannot be removed without breaking gauge
invariance. The last term is also a local polynomial that breaks
parity but it can be removed preserving gauge invariance. This term
illustrates that the $\zeta$-function prescription may yield
unessential pseudo-parity even contributions to the anomaly; the same
holds for chiral symmetry in even dimensions. The (essential) parity
anomaly is thus
\begin{equation}
\A_P= \eta\sigma\int dx_0\tr(A_0)= \eta\sigma W_{\rm CS}\,,
\end{equation}
where $W_{\rm CS}$ is the 0+1-dimensional version of the
Chern-Simons action. This checks that the essential parity anomaly is
a local polynomial and temperature independent. When the gauge field
configurations are traceless, $W_{\rm CS}$ vanishes identically. This
is consistent with the fact that in this case the gauge group must be
a subgroup of SU$(N_f)$ which is simply connected and does not have
large gauge transformations in 0+1 dimensions. (The gauge group itself
may have large gauge transformations but the effective action must be
strictly gauge invariant since one could always deform the
configuration within SU$(N_f)$.)

The effective action can be written as the sum of components with well
defined pseudo-parity, namely, $W_\sigma^\pm(-A_0)=\pm
W_\sigma^\pm(A_0)$.  After removing the term $\sigma N_f\beta m/2$,
the pseudo-parity even component is gauge and parity invariant. The
odd component can be identified as the anomalous part of the effective
action at finite temperature in 0+1 dimensions since it contains all
the anomalies
\begin{equation}
W_{\rm anom}(\D)= W_\sigma^-(\D)= \frac{\eta}{4}\int dx_0\tr(\Phi_\sigma)\,,
\end{equation}
where
\begin{eqnarray}
\Phi_\sigma &=&
\frac{2}{\beta}\Omega^\prime_\sigma(\frac{\beta}{2}(m-A_0),0)-
\hbox{p.p.c.}  \nonumber \\ &=&
\frac{2}{\beta}\left[\log\left(e^{-\sigma\beta(m-A_0)}+1\right)
-\log\left(e^{-\sigma\beta(m+A_0)}+1\right)\right]\,.
\label{eq:new.39}
\end{eqnarray}
In this expression, ${\rm p.p.c.}$ stands for pseudo-parity conjugate,
$A_0\to -A_0$. Besides, the branch cut of the logarithm is to be taken
along the negative real axis to obtain the manifolds $\M_\sigma$ for
each $\sigma$. Alternatively $W_{\rm anom}(\D)$ can also be written as
\begin{equation}
W_{\rm anom}(\D)=\eta(\frac{\sigma}{2} W_{\rm CS} +\Gamma_{\rm odd})
\end{equation}
with
\begin{equation}
\Gamma_{\rm odd}= -\tr\left[\tanh^{-1}\left(\tanh\left(\frac{\beta
m}{2}\right)\tanh\left(\frac{\beta A_0}{2}\right)\right)\right]\,.
\end{equation}
This same function appears also in exact 2+1-dimensional results
\cite{Fosco:1997ei} (cf. subsection \ref{subsec:3.d}).

The anomalous action $W_{\rm anom}(\D)$ resulting from the calculation is
to be compared with the naive temperature independent action
\begin{equation}
\frac{1}{2}\eta\left(\sigma-\varepsilon(m)\right)W_{\rm CS}\,,
\end{equation}
which also saturates the parity and gauge variations of the full
effective action but displays a singularity at $m=0$ which is spurious
at finite temperature. In fact this is the zero temperature limit of
$W_{\rm anom}$. On the other hand, the perturbative result follows
from using
\begin{equation}
\Phi_\sigma= \frac{4\sigma}{e^{\sigma\beta m}+1}A_0+O(A_0^3) \,.
\label{eq:new.42}
\end{equation}
Substituting in $W_{\rm anom}$ yields
\begin{equation}
W_{\rm anom}(\D) = \frac{1}{2}\eta\left(\sigma-\tanh(\frac{\beta
 m}{2})\right) W_{\rm CS} +O(A_0^3)\,.
\label{eq:new.22}
\end{equation}
This is the usual perturbative result which violates gauge invariance
under homotopically non trivial gauge transformations.

It is noteworthy that in 0+1 dimensions the exact result can also be
written in closed form without gauge fixing
\begin{equation}
W_\sigma(\D)= -\tr\log(e^{-\sigma\beta m}\Omega^{\eta\sigma}+1)\,,\quad 
\Omega= Te^{-\int_0^\beta A_0(x_0)dx_0}\,.
\end{equation}

\section{Summary of the main results and discussion}
\label{sec:summary}

In this section we will summarize the results for the
2+1-dimensional case. The details of the calculation are given in
next sections. The calculation proceeds by: (i) Expanding the
expression in eq.~(\ref{eq:7}) to select terms with at most two spatial
covariant derivatives. (ii) Taking the Dirac trace. To make this step
simple, we assume the scalar field $M(x)$ to be a constant c-number
mass $m$. (iii) Moving the operators $D_i$ to the right. This generates
gauge covariant objects of the form $[D_i,X]$. (iv) Carrying out the
$\bfk$ integration. This kills all remaining gauge non invariant
objects $\tr\langle 0|XD_i|0\rangle$. This step requires to
diagonalize $D_0$ in the Hilbert space of flavor and time (with $\bfx$
fixed) and can be done in an $A_0$-stationary gauge. (v) Summing over
fermionic frequencies $\omega_n$. And (vi) integrating over $z$ in
closed form whenever possible.

$W_0(\D)$ denotes the zeroth order term in the expansion in powers of
$D_i$, $W_2^\pm(\D)$ denote the pseudo-parity even and odd second
order terms, respectively. The dependence on $\sigma$ and $\beta$ will
not be displayed explicitly.

\begin{equation}
W_0(\D) = \frac{1}{4\pi}\int d^3x\,\tr \left[ 
\left(\frac{2}{\beta}\right)^2m\phi_1(\frac{\beta}{2}(m-A_0))-
\left(\frac{2}{\beta}\right)^3\phi_2(\frac{\beta}{2}(m-A_0))
 -\frac{1}{3}\sigma m^3 \right] +{\rm p.p.c.}
\end{equation}
The functions $\phi_n(\omega)$ are of the form
$\partial_\omega^{-(n+1)}\tanh(\omega)$, that is, they are primitives
of $\tanh(\omega)$, with appropriate integration constants so that
they are even or odd functions of $\omega$. Their proper definition is
given in Appendix~\ref{app:B}. The function $\phi_0(\omega)$ has
already appeared in the 0+1-dimensional case. The notation
p.p.c. stands for pseudo-parity conjugate, $A_0\to -A_0$, and after
taking $\int d^3x\,\tr$ it coincides with the complex conjugate.

\begin{eqnarray}
W^+_2(\D) &=&
\frac{1}{4\pi}\int dx^3\,\tr
\Bigg[
\frac{1}{2}\left(\frac{2}{\beta}\right)^2
\phi_1(\frac{\beta}{2}(m-A_0))\A_i\frac{1}{\Dh_0}\A_i
-\frac{2}{\beta}\phi_0(\frac{\beta}{2}(m-A_0))
\A_i\left(\frac{1}{4}+\frac{1}{2}\frac{m}{\Dh_0}\right)\A_i
\nonumber \\ &&
+ \int_0^{-\sigma\infty}dt\,
	\tanh(\frac{\beta}{2}(m-t-A_0))
	\A_i\left(\frac{1}{4}\Dh_0+\frac{m^2}{\Dh_0}\right)
	\frac{1}{2(m-t)+\Dh_0}\A_i
\Bigg] + \hbox{p.p.c.}
\label{eq:new.84b}
\end{eqnarray}
The symbol $\Dh_0$ denotes the covariant derivative $\Dh_0(X)=
[D_0,X]$ and p.p.c. now includes $\Dh_0\to -\Dh_0$. In addition,
we have introduced the new field $\A_i(x)$ which is defined as any
solution (in the space of matrix valued functions, i.e., multiplicative
operators in $\bfx$ and $x_0$ spaces) of the equation
\begin{equation}
\Dh_0 E_i= \Dh_0^2\A_i.
\label{eq:new.80}
\end{equation}
Where $E_i(x)=F_{0i}=[D_0,D_i]$ is the electric field. Essentially,
$\A_i=\Dh_0^{-1}E_i$. However, since the operator $\Dh_0$ is singular,
$\Dh_0^{-1}E_i$ either does not exist or is not unique. Because
$\A_i$ is not unique, we will have to check that the final result is
independent of the particular solution taken. This check is done
below. Clearly, the space of solutions of this equation is gauge and
parity invariant.

\begin{eqnarray}
W_2^-(\D) &=&
\frac{i\eta}{8\pi}\int d^3x\epsilon_{ij}\tr\Bigg[
(\frac{1}{2}F_{ij}-\A_i\A_j)\Phi_\sigma
-\sigma\A_i\Dh_0\A_j \nonumber\\
&&
+2m\int_0^{-\sigma\infty} dt \A_i\left(
\frac{\tanh(\frac{\beta}{2}(m-t+A_0))}{2(m-t)+\Dh_0} -
\frac{\tanh(\frac{\beta}{2}(m-t-A_0))}{2(m-t)-\Dh_0}
\right)\A_j
\Bigg]\,.
\label{eq:8.b}
\end{eqnarray}
The quantity $\Phi_\sigma$ was introduced in eq.~(\ref{eq:new.39})
and $F_{ij}=[D_i,D_j]$.

As already noted at the end of subsection~\ref{subsec:2.g}, a further
expansion can be done in powers of $\Dh_0$. This expansion can be
regarded as one in powers of $\Dh_\mu=[D_\mu,\,]$, since $D_i$
appears only in commutators. Such an expansion preserves two essential
properties of the original expansion in powers of $D_i$: first, it
does not break gauge or parity symmetries and second, higher order
terms are increasingly ultraviolet convergent.

Clearly, $W_0$ is already independent of $\Dh_0$. The expansion of
$W_2^\pm$ is less obvious and is done in their respective sections.
For the pseudo-parity even component we find
(cf. subsection~\ref{subsec:new.2.f}) 
\begin{equation}
W^+_{2,\,{\rm leading}}(\D)=
-\frac{1}{24\pi}\int d^3x\,\tr\,\Big[
\frac{\tanh(\frac{\beta}{2}(m-A_0))}{2m}
-\frac{\beta}{2}\frac{1}{4\cosh^2(\frac{\beta}{2}(m-A_0))}
\Big]E_i^2 + \hbox{p.p.c.}
\label{eq:new.86b}
\end{equation}
On the other hand, for the pseudo-parity odd sector, the leading order
in $\Dh_0$ is (cf. section~\ref{sec:4}):
\begin{eqnarray}
W_{\rm anom}(\D) &=& \frac{i\eta}{8\pi}\int d^3x \epsilon_{ij}\tr\Bigg[
(\frac{1}{2}F_{ij}-\A_i\A_j)\Phi_\sigma
\nonumber\\ && \quad
+\frac{1}{2}\Big(\tanh(\frac{\beta}{2}(m-A_0))+
\tanh(\frac{\beta}{2}(m+A_0))-2\sigma\Big)\A_i\Dh_0\A_j \Bigg]\,.
\label{eq:22b}
\end{eqnarray}
This component of the action differs from $W^-_2$ by terms which are
ultraviolet convergent, gauge and parity invariant and free of
multivaluation, thus it is natural to identify it with the purely
anomalous component of the effective action: it is the ultraviolet
logarithmically divergent part of the effective action that contains
all the essential anomaly.

In the remainder of this section we will discuss the properties of
$W_{\rm anom}(\D)$ since this term has attracted more attention in the
literature. The discussion of $W_0$ and $W^\pm_2$ will be made in
their respective sections. Nevertheless, it can be pointed out that
the pseudo-parity even components are strictly gauge invariant, not
just modulo $2\pi i$. This was to be expected since the only allowed
ambiguity in the $\zeta$-function regularized action is $2\pi in$
which is purely imaginary and $W^+$ is real. Also they are parity
invariant, except for the term $-\frac{1}{3}\sigma m^3$ in $W_0$ which
is is removable by counterterms. On the other hand, $W^-_2$ has the
same parity and gauge transformations as $W_{\rm anom}(\D)$.  No scale
anomaly is present in the effective action.

First of all, let us check that $W_{\rm anom}(\D)$ does not depend on the
particular solution taken for $\A_i(x)$. In order to show this, note
first that if $A_0(\bfx)$ is not only stationary but also diagonal
$[A_0(\bfx),\partial_iA_0(\bfx)]=0$, and a solution is simply
$\A_i=A_i$. In an arbitrary $A_0$-stationary gauge, an explicit
solution is $\A_i=A_i-U\partial_iU^{-1}$, where $U(\bfx)$ is any of
the stationary gauge transformations bringing $A_0(\bfx)$ to diagonal
form. Adding an arbitrary solution of $\Dh_0(X)=0$, the most general
form of $\A_i$ is
\begin{equation}
\A_i(x)=A_i(x)-U(\bfx)\partial_iU^{-1}(\bfx) + B_i(\bfx)\,, \quad
[A_0(\bfx),B_i(\bfx)]=0\,.
\label{eq:11}
\end{equation}
$B_i(\bfx)$ is time-independent and commutes with $A_0(\bfx)$ at each
point and is otherwise arbitrary.\footnote{This is the generic
case. If two different eigenvalues of $A_0(\bfx)$ differ by $2\pi
in/\beta$ in a sufficiently large region of the space-time,
$\Dh_0(X)=0$ will admit time-dependent and not $A_0$-commuting
solutions as well.} An explicit calculation shows that, as expected,
the field $\A_i$ transforms covariantly under stationary and discrete
gauge transformations, modulo redefinitions of $B_i$. It is invariant
under parity transformations. Using antisymmetry of $\epsilon_{ij}$
and the cyclic property of the trace, it easy to see that $W_{\rm
anom}(\D)$ is uniquely defined, that is, the dependence on $B_i(\bfx)$
cancels in eq.~(\ref{eq:22b}). We remark that $A_i$ depends both on
$\bfx$ and $x_0$, in general.

\subsection{Parity variation}
\label{subsec:3.b}
Let us study the properties of $W_{\rm anom}(\D)$ under parity
transformations. As expected, replacing $\D$ by $\D^P$ and $\sigma$ by
$-\sigma$ leaves the formula unchanged thus eq.~(\ref{eq:18b}) is
verified. Therefore, the anomaly can be obtained from the
relation $\A_P(\D)= W_\sigma(\D)-W_{-\sigma}(\D)$. Using the property
\begin{equation}
\Phi_\sigma-\Phi_{-\sigma} = 4\sigma A_0\,,\quad\hbox{on $\Mt$}\,,
\end{equation}
and substituting in $W_{\rm anom}(\D)$, the parity anomaly is
\begin{equation}
\A_P(\D)=
\eta\sigma\frac{i}{4\pi}\int d^3x\, \epsilon_{ij}\tr\Bigg[
(F_{ij}-2\A_i\A_j) A_0-\A_i\Dh_0\A_j \Bigg]\,.
\end{equation}
This expression can be simplified by using the following replacement
(valid inside $\int d^3x\,\tr$)
\begin{equation}
\epsilon_{ij}\A_i\Dh_0\A_j= \epsilon_{ij}(A_i\partial_0 A_j-2\A_i\A_j A_0)\,.
\label{eq:new.20}
\end{equation}
This follows from noting that $\A_i\partial_0\A_j$ can be replaced by
$A_i\partial_0A_j$ since $\A_i$ and $A_i$ differ only by a stationary
field (cf. eq.~(\ref{eq:11})) and integration by parts. Then, the
terms containing $\A_i$ cancel and the final expression for the parity
anomaly becomes
\begin{equation}
\A_P(\D)=\eta\sigma\frac{i}{4\pi}\int
d^3x\epsilon_{ij}\tr(A_0F_{ij} - A_i\partial_0A_j)
=\eta\sigma W_{\rm CS}(\D).
\end{equation}
$W_{\rm CS}$ being the Chern-Simons action. This checks that the
parity anomaly has a temperature and mass independent form and is a
local polynomial, as expected from general arguments given in
subsection~\ref{subsec:2.d}.

\subsection{Gauge variation}
Any correct calculation must be consistent with gauge invariance,
however, since we have partially fixed the gauge, we can only check
invariance under the restricted set of transformations considered in
subsection~\ref{subsec:2.c}, i.e., time-independent and discrete gauge
transformations. In other words, it must be verified that bringing the
gauge configuration to two different $A_0$-stationary gauges gives the
same result modulo $2\pi i$. As has been noted in
subsection~\ref{subsec:2.c}, for some gauge configurations all allowed
gauge transformations (i.e., consistent with the condition
$\partial_0A_0=0$) are topologically small and on the other hand,
going to a $A_0$-stationary gauge may require large
transformations. Thus our calculation does not constitute a proof that
the effective action is invariant under all gauge
transformations. This fact has already been established as a
consequence of using the $\zeta$-function renormalization prescription
and the fact that gauge transformations do not change the spectrum of
the Dirac operator.

In order to study the gauge variation of $W_{\rm anom}(\D)$, note that
all the quantities there transform covariantly under time-independent
gauge transformations. In particular, $\A_i$ is defined by a covariant
equation. Regarding discrete gauge transformations, $A_0$ is the only
quantity which transforms inhomogenously, namely
$A_0^U(\bfx)=A_0({\bfx})+\Lambda(\bfx)$. The terms with $\tanh$ in
$W_{\rm anom}(\D)$ depend on $\exp(\beta A_0)$ and this quantity is
unchanged by discrete gauge transformations, cf. eq~(\ref{eq:A2}),
hence the last term in $W_{\rm anom}(\D)$ is invariant\footnote{It is
interesting to note that the two tanh terms in $W_{\rm anom}$ can be
combined into $\frac{\sinh(\beta m)}{\cosh(\beta m)+\cosh(\beta A_0)}$
which is the same combination appearing in \cite{Sissakian:1997cp}
replacing $A_0$ by the chemical potential.}. The only variation may
come through $\Phi_\sigma$. This quantity also depends on $\exp(\beta
A_0)$, thus if the argument of the logarithm is chosen always on the
same Riemann sheet, i.e., on the cut manifold $\M_\sigma$, this
quantity is trivially invariant. On the other hand, on the Riemann
surface $\Mt$ one can consider a path connecting $\D$ and $\D^U$, for
instance the path $A_0^t=A_0+t\Lambda$ with $0\le t\le 1$ and keeping
$m$ fixed. Let us consider first the generic non-degenerated case
discussed in subsection~\ref{subsec:2.c} for which eqs.~(\ref{eq:A3})
apply, then a simple analysis shows that
\begin{equation}
\Phi_\sigma^U-\Phi_\sigma= 4\sigma\Theta(-\sigma m)\Lambda\,,
\quad\hbox{on $\Mt$}\,.
\end{equation}
This implies that the corresponding gauge variation of $W_{\rm
anom}(\D)$ is of the form $(\sigma-\varepsilon(m))$ times something
independent of $\sigma$. (Here $\Theta$ and $\varepsilon$ stand for
the step and sign functions, respectively.) Because it has already
been established in the previous subsection that the
$\sigma$-dependent part of the action is $\frac{1}{2}\eta\sigma W_{\rm
CS}$, it follows that $W_{\rm anom}(\D)$ must have precisely the same
gauge variation as $\frac{1}{2}\eta(\sigma-\varepsilon(m)) W_{\rm
CS}$, that is
\begin{equation}
W_{\rm anom}(\D^U)-W_{\rm anom}(\D) = 
\eta\sigma\Theta(-\sigma m)(W_{\rm CS}(\D^U)-W_{\rm CS}(\D))\,,
\quad\hbox{on $\Mt$}\,.
\end{equation}
Observe that when $\sigma m>0$ the formula predicts strict gauge
invariance. This is correct since in this case the branch cut $\Gamma$
and the spectrum do not intersect and so there is no flux of
eigenvalues through the branch cut when going from $\D$ to $\D^U$. As
noted previously, the variation of the Chern-Simons action is $2\pi i$
times an integer which depends on the topological numbers of the gauge
transformation and the gauge field configuration. Then, as expected,
both $W_{\rm anom}$ and $\A_P$ are gauge invariant modulo $2\pi i$.

For degenerated configurations (defined in subsection~\ref{subsec:2.c}),
$\Lambda$ no longer needs to commute with $A_0$, however, by
inspection of $\Phi_\sigma$ and unitarity of $\exp(\beta A_0)$, it
follows that there will be no jump in the logarithm unless $\sigma m
<0$ and in this case the jump is odd in $\sigma$, therefore the same
factor $\sigma\Theta(-\sigma m)$ is obtained and the same argument as
before applies.

\subsection{Abelian-like reductions}
\label{subsec:3.d}
As a check of our formula, let us assume that $\bfA$ is also
time-independent and commutes everywhere with $A_0$. In this case, the
field $\A_i$ has (or can be chosen to have) the same properties. As a
consequence all terms with $\A_i$ vanish due to antisymmetry and our
result for $W_{\rm anom}(\D)$ reduces to
\begin{equation}
W_{\rm anom}(\D) = \frac{i\eta}{16\pi}\int
d^3x\epsilon_{ij}\tr\left(F_{ij}\Phi_\sigma\right)\,.
\label{eq:new.21}
\end{equation}
This expression holds also for $W^-_2$. Separating the even and odd
components under $m\to -m$, this can be rewritten as
\begin{equation}
W_{\rm anom}(\D)=\eta(\frac{\sigma}{2} W_{\rm CS} +\Gamma_{\rm odd}) \,,
\end{equation}
 where
\begin{equation}
\Gamma_{\rm odd}= -\frac{i}{4\pi}\int
d^2x\epsilon_{ij}\tr\left[F_{ij}\tanh^{-1}\left(\tanh\left(\frac{\beta
m}{2}\right)\tanh\left(\frac{\beta A_0}{2}\right)\right)\right]
\end{equation}
is the expression introduced in Ref.\cite{Fosco:1997ei}. As proven
there (see also
\cite{Deser:1997nv,Fosco:1998cq,Felipe:1997nx}),
when $A_0$ is in addition $\bfx$-independent, there are no higher
order corrections and the right-hand side of eq.~(\ref{eq:new.21})
gives the full pseudo-parity odd component of the effective
action. (Recall that according to our definitions, $\Gamma_{\rm odd}$
is even under parity since $m$ and $A_0$ are both odd. $\Gamma_{\rm
odd}$ is an odd function of $m$.)

\subsection{Zero temperature limit}
Using the zero temperature limit of $\Phi_\sigma$,
\begin{equation}
\Phi_\sigma= 4\sigma\Theta(-\sigma m)A_0, \quad (T=0)\,,
\end{equation}
and the replacement in eq.~(\ref{eq:new.20}), it is straightforward to
derive the zero temperature limit of our formula, namely
\begin{equation}
W_{\rm anom}(\D)= \frac{1}{2}\eta\left(\sigma-\varepsilon(m)\right)
 W_{\rm CS}\,, \quad(T=0)\,.
\label{eq:19}
\end{equation}
This is the standard result at zero temperature (see e.g.
\cite{GamboaSaravi:1996aq,Salcedo:1996qy}).

\subsection{Massless fermions}
We may also consider massless fermions. In this case the terms with
$\tanh$ cancel in $W_{\rm anom}(\D)$. Also, $\Phi_\sigma$ becomes an
odd function of $\sigma$ on $\Mt$. Since all terms in the action are
odd under $\sigma\to -\sigma$, it follows from subsection~\ref{subsec:3.b} that
\begin{equation}
W_{\rm anom}(\D)=
\frac{1}{2}\eta\sigma W_{\rm CS}\,,\quad\hbox{for $m=0$ and on
$\Mt$}\,.
\label{eq:new.23}
\end{equation}
This formula holds for $W^-_2$ too. This refers to the formula after
analytical extension, i.e., on $\Mt$. Keeping track of the Riemann
sheet in the logarithm in $\Phi_\sigma$ adds a term $i\pi n$, so that
the trivial gauge invariance of the $\zeta$-function prescription is
preserved on $\M_\sigma$. The full result is an application of the
Atiyah-Singer index
theorem\cite{Alvarez-Gaume:1985nf,Deser:1997nv}. Because in the
massless case parity and pseudo-parity transformations coincide,
$W^-(\D)$ equals $\frac{1}{2}\A_P(\D)$. Thus, eq.~(\ref{eq:new.23}) is
in fact exact to all orders\cite{Alvarez-Gaume:1985nf}.

\subsection{Perturbative result}

Next, let us show that our result for the pseudo-parity odd part
reproduces the results obtained using perturbation theory at lowest
orders. As noted in the Introduction, the latter approach yields a
renormalization factor in front of the Chern-Simons action which is
not quantized, and so breaks gauge invariance under large gauge
transformations. To obtain the perturbative result we should retain terms of
zeroth or first order in $A_0$ in $W_{\rm anom}(\D)$. Use of the
replacement in eq.~(\ref{eq:new.20}) and the expansion of
$\Phi_\sigma$ in eq.~(\ref{eq:new.42}), gives
\begin{eqnarray}
W_{\rm anom}(\D) &=& \frac{i\eta}{8\pi}\int d^3x \epsilon_{ij}
\tr\Bigg[ \frac{4\sigma}{e^{\sigma\beta
m}+1}(\frac{1}{2}F_{ij}-\A_i\A_j)A_0 \nonumber\\ && \quad
+\left(\tanh(\frac{\beta m}{2})-\sigma\right)( A_i\partial_0
A_i-2\A_i\A_j A_0) \Bigg] +O(A_0^3)\,.
\end{eqnarray}
It trivial to check that the terms with $\A_i$ cancel, and the final
result can be expressed as
\begin{equation}
W_{\rm anom}(\D) = \frac{1}{2}\eta\left(\sigma-\tanh(\frac{\beta
 m}{2})\right) W_{\rm CS} +O(A_0^3)\,,
\end{equation}
which is the standard perturbative
result\cite{Ishikawa:1987zi}. It has the same form as in
the 0+1-dimensional case, eq.~(\ref{eq:new.22}).

\subsection{Relation to the chiral case}
There are strong similarities with the situation of chiral anomalies
in even dimensions. There the (consistent) chiral anomaly is defined
as the variation of the effective action under chiral
transformations. The chiral anomaly contains an essential part which
can only be derived from a non polynomial action. As it is well-known,
at zero temperature this is the gauged WZW action which again is
pseudo-parity odd\cite{Witten:1983tw}. As already noted, the chiral
anomaly has also been shown to be temperature independent (see
e.g.\cite{GomezNicola:1994vq}). This observation has lead to propose
that the same gauged WZW action is the full anomalous action at finite
temperature too\cite{Alvarez-Estrada:1993jm}. The findings in 0+1 and 2+1
dimensions suggest that this is not the case. Indeed, from the point
of view of parity anomaly saturation and gauge invariance, the naive
result given by eq.~(\ref{eq:19}) is entirely sufficient, however the
correct result, eq.~(\ref{eq:22b}), has an explicit temperature
dependence. It is worth noting that the naive formula has a
singularity along the line $m=0$ which is spurious at finite
temperature, since there are no infrared singularities, and is not
present in the full result. If a similar situation takes place in the
chiral case, the amplitude corresponding to anomalous processes (but
not the anomaly itself) will show a smooth temperature dependence. (By
anomalous processes it is meant those processes driven by a
logarithmically divergent pseudo-parity odd amplitude.) Such
dependence has actually been found in Ref.\cite{Pisarski:1996ne}. A
study of the chiral case along the lines followed here has been
carried out in\cite{Salcedo:1998tg}, with the result that the
anomalous component of the effective action has indeed a non trivial
temperature dependence.

\section{The effective action at zeroth order}
\label{sec:new.1}

In this section we will compute the 2+1-dimensional effective action
at lowest order in the derivative expansion. For simplicity we will
use units $\beta=2$. After setting $D_i$ to zero, eq.~(\ref{eq:7})
yields\footnote{The subindex 0 in $\Omega_{s,0}(\D)$ stands for zeroth
order. We will no longer make explicit the dependence on $\Gamma$.}
\begin{equation}
\Omega_{s,0}(\D)= -\int\frac{d^2k}{(2\pi)^2}\frac{1}{2}\sum_n\int_\Gamma
\frac{dz}{2\pi i}z^s\tr\langle 0|\frac{1}{\gamma_0 Q+\bfgamma\bfp+\mu}
|0\rangle
\,.
\end{equation}
Here $Q= \omega_n+D_0$ and $\mu=M(x)-z$. As mentioned, we will take
the scalar field $M(x)=m$. Then, $Q$, $\bfp$ and $\mu$ commute with
each other and $\gamma_0 Q+\bfgamma\bfp$ can be brought to the
numerator
\begin{equation}
\frac{1}{\gamma_0 Q+\bfgamma\bfp+\mu}= 
\frac{\mu-\gamma_0Q-\bfgamma\bfp}{\Delta}\,,
\quad \mu=m-z\,,\quad \Delta= \mu^2-Q^2+k^2\,.
\label{eq:new.3}
\end{equation}
The Dirac trace can be computed immediately yielding
\begin{equation}
\Omega_{s,0}(\D)= -\int\frac{d^2k}{(2\pi)^2}\sum_n\int_\Gamma
\frac{dz}{2\pi i}z^s\tr\langle 0| \frac{\mu}{\Delta} |0\rangle \,.
\end{equation}
Here the trace refers to flavor only. In order to perform the momentum
integrals it is convenient to use integration by parts:
\begin{equation}
-\int_\Gamma\frac{dz}{2\pi i}z^s f(z) = 
\frac{1}{s+1}\int_\Gamma\frac{dz}{2\pi i}z^{s+1}f^\prime(z)\,,
\label{eq:39}
\end{equation}
which holds provided that $z^sf(z)$ vanishes at infinite on $\Gamma$
for sufficiently negative ${\rm Re}(s)$. Then
\begin{eqnarray}
\Omega_{s,0}(\D) &=& -\int\frac{d^2k}{(2\pi)^2}\sum_n\int_\Gamma
\frac{dz}{2\pi i}z^s\tr\langle 0| 
\left(
-\frac{z^2}{(s+1)(s+2)}\frac{2\mu}{\Delta^2}
-\frac{z}{(s+1)}\frac{2\mu^2}{\Delta^2}
\right)
 |0\rangle 
\nonumber \\
&=& -\int\frac{d^2k}{(2\pi)^2}\sum_n\int_\Gamma
\frac{dz}{2\pi i}z^{s+1}\tr\langle 0| 
\left(\frac{z}{s+2}-\frac{m}{s+1}\right)\frac{2\mu}{\Delta^2}
|0\rangle \,.
\end{eqnarray}
The momentum integral can be done using
\begin{equation}
\int\frac{d^2k}{(2\pi)^2}\frac{k^{2n}}{(k^2+M^2)^N} =
\frac{\Gamma(n+1)\Gamma(N-n-1)}{4\pi\Gamma(N)}(M^2)^{n-N+1}\,
\label{eq:new.4}
\end{equation}
and it yields
\begin{equation}
\Omega_{s,0}(\D)= -\frac{1}{4\pi}\sum_n\int_\Gamma
\frac{dz}{2\pi i}z^{s+1}\tr\langle 0| 
\left(\frac{z}{s+2}-\frac{m}{s+1}\right)
\frac{2\mu}{\mu^2-Q^2}
|0\rangle \,.
\end{equation}
The sum over frequencies can be done straightforwardly by using
eq.~(\ref{eq:new.1})
\begin{equation}
\Omega_{s,0}(\D) =
 -\frac{1}{4\pi}\int d^3x\int_\Gamma
\frac{dz}{2\pi i}z^{s+1}\tr
\left(\frac{z}{s+2}-\frac{m}{s+1}\right)
\tanh(\mu-A_0)
+{\rm p.p.c.}
\end{equation}
Again, p.p.c. stands for pseudo-parity conjugate. In addition, we have
used that $A_0$ is time independent and so $D_0$ can be replaced by
$A_0$.

Finally, the integration over $z$ can be done using the functions
$\Omega_\Gamma(\omega,s)$ introduced in eq.~(\ref{eq:new.7}) and
Appendix~\ref{app:B}. Thus
\begin{equation}
\Omega_{s,0}(\D) =
 \frac{1}{4\pi}\int d^3x\,\tr
\left(\frac{1}{s+2}\Omega_\Gamma(m-A_0,s+2)
-\frac{m}{s+1}\Omega_\Gamma(m-A_0,s+1)
\right)
+{\rm p.p.c.}
\label{eq:new.50}
\end{equation}
To obtain the effective action it remains to apply
$-\frac{d}{ds}|_{s=0}$. As shown in Appendix~\ref{app:B}
\begin{eqnarray}
\Omega_\sigma(\omega,n) &=& 0\,,\quad n=0,1,2,\dots\,, \nonumber \\
\Omega^\prime_\sigma(\omega,1) &=& \phi_1(\omega)-
\sigma\left(\frac{1}{2}\omega^2-\frac{1}{6}\left(\frac{i\pi}{2}\right)^2\right)
\,, \label{eq:new.51} \\
\frac{1}{2!}\Omega^\prime_\sigma(\omega,2) &=& \phi_2(\omega)-
\sigma\left(\frac{1}{6}\omega^3- 
\frac{1}{6}\left(\frac{i\pi}{2}\right)^2\omega\right) 
\,.  \nonumber
\end{eqnarray}
Here the prime refers to derivative with respect to $s$ and
$\sigma=\pm 1$ refers to the two possible inequivalent choices of
$\Gamma$. The functions $\phi_n(\omega)$ are defined for every integer
$n$ and satisfy
\begin{eqnarray}
\phi_{-1}(\omega) &=&\tanh(\omega)\,,  \label{eq:new.52.a} \\
\phi_n^\prime(\omega) &=& \phi_{n-1}(\omega)\,, \label{eq:new.52.b}\\
\phi_n(-\omega) &=& (-1)^n\phi_n(\omega) \label{eq:new.52.c} \,.
\end{eqnarray}
(See Appendix~\ref{app:B} for the fixing of the integration constants
for non negative $n$.) The function $\phi_0(\omega)$ has already
appeared in the 0+1-dimensional case and similar comments can be
made here: for non negative integer $n$, the terms $\phi_n(\omega)$ in
$\Omega^\prime_\Gamma(\omega,n)$ are those preserving parity and the
breaking comes from the polynomial term which is odd under $\sigma\to
-\sigma$.

After taking the $-\frac{d}{ds}|_{s=0}$ in eq.~(\ref{eq:new.50}) and using
eqs.~(\ref{eq:new.51}), the effective action at zeroth order takes the
form
\begin{equation}
W_0(\D) = \frac{1}{4\pi}\int d^3x\,\tr \left(
m\phi_1(m-A_0)-\phi_2(m-A_0) -\frac{1}{3}\sigma m^3 \right) +{\rm
p.p.c.}
\end{equation}
Recalling that $m$ and $A_0$ change sign under parity, it follows that
this action preserves parity except for the last term $\sigma m^3$,
which is similar to the term $\sigma m$ in 0+1 dimensions. Since
this term is polynomial and strictly gauge invariant it can be removed
from the action.

Next we should check the gauge invariance of $W_0(\D)$. Under
time-independent gauge transformations $A_0$ changes by a similarity
transformation and this leaves $W_0(\D)$ invariant due to the cyclic
property of the trace. Under discrete gauge transformations the
spectrum of $A_0$ changes by an integer multiple of $i\pi$. The strict
gauge invariance of $\Omega_{s,0}(\D)$, and hence of $W_0(\D)$, on the
manifold $\M_\sigma$ (i.e., the complex plane cut along $\Gamma$)
follows immediately from eqs.~(\ref{eq:new.7}) and (\ref{eq:new.50})
since $\tanh$ is periodic and so the functions
$\Omega_\Gamma(\omega,s)$ are also periodic. The strict gauge
invariance of $W_0(\D)$ also on the Riemann surface $\Mt$ can be
checked directly using the formula (cf. Appendix~\ref{app:B})
\begin{equation}
\phi_n(\omega+\frac{i\pi}{2})-\phi_n(\omega-\frac{i\pi}{2})
=i\pi\varepsilon(m)\frac{\omega^n}{n!}\,,\quad n=0,1,2,\dots
\label{eq:new.54}
\end{equation}
where we have used the notation $m={\rm Re}(\omega)$ (since in
practice $\omega=m\mp A_0$ and $A_0$ is anti-Hermitian),
$\varepsilon(x)$ is the sign of $x$, and the difference refers to the
straight path from $\omega-\frac{i\pi}{2}$ to $\omega+\frac{i\pi}{2}$
on the complex plane. This identity can be used to show that the
combination appearing in $W_0(\D)$,
$m\phi_1(\omega)-\phi_2(\omega)+{\rm c.c.}$ (where c.c. stands for
complex conjugate) is a periodic function.

\section{The pseudo-parity even effective action at second order}
\label{sec:new.2}

\subsection{Dirac degrees of freedom}

Starting from eq.~(\ref{eq:7}), the second order takes the form (we
use units $\beta=2$)
\begin{equation}
\Omega_{s,2}(\D)= -\int\frac{d^2k}{(2\pi)^2}\frac{1}{2}\sum_n\int_\Gamma
\frac{dz}{2\pi i}z^s\tr\langle 0|\left(\frac{1}{\gamma_0 Q+\bfgamma\bfp+\mu}
\bfgamma \bfD\right)^2\frac{1}{\gamma_0 Q+\bfgamma\bfp+\mu}
|0\rangle
\,.
\label{eq:new.57}
\end{equation}
Where $Q$ and $\mu$ were introduced in the previous section. Using
again eq.~(\ref{eq:new.3}), the Dirac trace can be computed
immediately.  Since we are interested in the pseudo-parity even
sector, $W_2^+(\D)$, we keep only terms without the Levi-Civita
pseudo-tensor $\epsilon_{ij}$. The resulting expression is simplified
using that $k_ik_j$ is equivalent to $\frac{1}{2}k^2\delta_{ij}$
within the two-dimensional momentum integral. This yields
\begin{eqnarray}
\Omega_{s,2}^+(\D) &=& -\int\frac{d^2k}{(2\pi)^2}\sum_n\int_\Gamma
\frac{dz}{2\pi i}z^s \mu\, \tr\langle 0|\left(
	(\mu^2-k^2)\frac{1}{\Delta}D_i\frac{1}{\Delta}D_i\frac{1}{\Delta}
\right.
\nonumber\\
&&
\left.
	+\frac{Q}{\Delta}D_i\frac{1}{\Delta}D_i\frac{Q}{\Delta}
	-\frac{1}{\Delta}D_i\frac{Q}{\Delta}D_i\frac{Q}{\Delta}
	-\frac{Q}{\Delta}D_i\frac{Q}{\Delta}D_i\frac{1}{\Delta}
\right)|0\rangle
\,.
\label{eq:new.58}
\end{eqnarray}
Here the trace refers to flavor only. 

\subsection{Space-time and flavor degrees of freedom}

Next, the spatial covariant derivatives are brought to the right
producing the quantity $E_i=[Q,D_i]=F_{0i}$. We have integrated by
parts terms of the form $[D_i,E_i]$ in order to produce a simpler
expression. In this form the following expression is derived:
\begin{eqnarray}
\Omega_{s,2}^+(\D) &=& 
-\int\frac{d^2k}{(2\pi)^2}\sum_n\int_\Gamma
\frac{dz}{2\pi i}z^s \mu\,\tr\langle 0|
\Bigg[
(\mu^2-k^2-Q^2)\frac{1}{\Delta^3}D_i^2
\nonumber\\ &&
+\Bigg(
\frac{1}{\Delta}E_i\frac{Q}{\Delta^2}
+\frac{Q}{\Delta}E_i\frac{1}{\Delta^2}
+\frac{Q^3}{\Delta^2}E_i\frac{1}{\Delta^2}
+\frac{Q^2}{\Delta^2}E_i\frac{Q}{\Delta^2}
+2\frac{Q^3}{\Delta^3}E_i\frac{1}{\Delta}
+2\frac{Q^2}{\Delta^3}E_i\frac{Q}{\Delta}
\nonumber\\ &&
-(\mu^2-k^2)\left(
\frac{1}{\Delta^2}E_i\frac{Q}{\Delta^2}
+\frac{Q}{\Delta^2}E_i\frac{1}{\Delta^2}
+2\frac{Q}{\Delta^3}E_i\frac{1}{\Delta}
+2\frac{1}{\Delta^3}E_i\frac{Q}{\Delta}
\right)
\Bigg)D_i
\nonumber\\ &&
-\frac{1}{\Delta}E_i\frac{1}{\Delta^2}E_i
+\frac{1}{\Delta^2}E_i\frac{Q^4}{\Delta^3}E_i
+\frac{Q^2}{\Delta^2}E_i\frac{Q^2}{\Delta^3}E_i
+2\frac{Q}{\Delta^2}E_i\frac{Q^3}{\Delta^3}E_i
\nonumber\\ &&
-(\mu^2-k^2)\left(
\frac{1}{\Delta^2}E_i\frac{Q^2}{\Delta^3}E_i
+\frac{1}{\Delta^3}E_i\frac{Q^2}{\Delta^2}E_i
+2\frac{Q}{\Delta^2}E_i\frac{Q}{\Delta^3}E_i
\right)
\Bigg]|0\rangle
\,.
\label{eq:new.43a}
\end{eqnarray}
Several remarks are in order here. First, the terms with explicit
$D_i$ break gauge invariance and will be shown to cancel after
momentum integration. That these terms break gauge invariance can be
seen from $D_i|0\rangle=A_i|0\rangle$. The breaking is due to the
state $|0\rangle$. The terms of the form $E_iE_i$ are explicitly
invariant under time-independent gauge transformations, since both
$E_i$ and $Q=D_0+\omega_n$ transform covariantly under such
transformations. For these terms, the operators appearing in $\langle
0||0\rangle$ are purely multiplicative in $\bfx$-space since all
$\partial_i$ operators appear inside commutators. Finally, to obtain a
more compact expression, the cyclic property has been used in the
gauge invariant terms. This is an important point of the present
formalism and deserves further clarification. The gauge invariant
terms contain the construction
\begin{equation}
\langle X\rangle = \tr\langle 0|X|0\rangle \,,
\end{equation}
where $X$ is purely multiplicative in $\bfx$-space. In this case, we
can consider $\langle X\rangle$ for each $\bfx$ separately and use
$|0\rangle$ to refer to the zero-frequency state in the Hilbert space
of functions of $x_0$. The point is that $\langle\,\rangle$ is not a
trace in $x_0$-space and thus the cyclic property holds only in a
restricted form. Nevertheless two cyclic properties can be used:

{\bf Rule 1.} {\em If $X(D_0)$ is a function of $D_0$ only and $Y$ is
a multiplicative operator in $\bfx$-space,
\begin{equation}
\langle XY\rangle =\langle YX\rangle \,.
\label{eq:cyclic.1}
\end{equation}
}

{\bf Rule 2.} {\em If $X_1(D_0)$ and $X_2(D_0)$ are functions of $D_0$
only, and $Y_1$ and $Y_2$ are multiplicative operators both in $\bfx$
and $x_0$ the following substitution applies
\begin{equation}
\sum_n\langle X_1(Q)Y_1X_2(Q)Y_2\rangle = 
\sum_n\langle X_2(Q)Y_2X_1(Q)Y_1\rangle \,,
\label{eq:cyclic.2}
\end{equation}
provided that the sum over $n$ is sufficiently convergent.
}(The sum refers to the $n$ dependence in $Q=D_0+\omega_n$.)

To see this, we exploit that $A_0$ is stationary to introduce the
basis of the flavor-time space (for given $\bfx$) formed with
eigenstates of $Q$, $|\alpha,\ell\rangle$, namely
\begin{eqnarray}
A_0|\alpha\rangle &=& a_\alpha|\alpha\rangle\,,\quad
\langle x_0|\ell\rangle =e^{\kappa_\ell
x_0}\quad(\kappa_\ell=i\pi\ell)\,,
\nonumber \\
|\alpha,\ell\rangle &=& |\alpha\rangle\otimes|\ell\rangle\,,
 \nonumber\\
Q|\alpha,\ell\rangle &=& (\omega_n+a_\alpha+\kappa_\ell)|\alpha,\ell\rangle\,.
\end{eqnarray}
Then, the first rule follows from
\begin{equation}
\langle X(D_0)Y\rangle = \sum_\alpha X(a_\alpha)
\langle\alpha,0|Y|\alpha,0\rangle
= \langle YX(D_0)\rangle\,.
\end{equation}
And the second rule comes from
\begin{eqnarray}
\sum_n\langle X_1(Q)Y_1 X_2(Q)Y_2\rangle &=&
\sum_n\sum_{\alpha,\beta}\sum_\ell
X_1(\omega_n+a_\alpha) \langle\alpha,0|Y_1|\beta,\ell\rangle 
X_2(\omega_n+a_\beta+\kappa_\ell) \langle\beta,\ell|Y_2|\alpha,0\rangle
\nonumber\\ =&&
\sum_n\sum_{\alpha,\beta}\sum_\ell
X_2(\omega_n+a_\alpha-\kappa_\ell) \langle\alpha,-\ell|Y_2|\beta,0\rangle
X_1(\omega_n+a_\beta) \langle\beta,0|Y_1|\alpha,-\ell\rangle 
\nonumber\\ =&&
\sum_n\sum_{\alpha,\beta}\sum_\ell
X_2(\omega_n+a_\alpha) \langle\alpha,0|Y_2|\beta,\ell\rangle
X_1(\omega_n+a_\beta+\kappa_\ell) \langle\beta,\ell|Y_1|\alpha,0\rangle 
\nonumber\\ =&&
\sum_n\langle X_2(Q)Y_2X_1(Q)Y_1\rangle\,.
\end{eqnarray}
The first equality uses that $Y_{1,2}$ are multiplicative in
$\bfx$-space.  The second equality comes from exchanging $\alpha$ with
$\beta$ and $\ell$ by $-\ell$. The third equality uses that $Y_{1,2}$
are multiplicative operators in $x_0$-space and also $n$ has been
shifted to $n+\ell$. This last step means that the formula only needs
to hold if $\sum_n\langle X_1(Q)Y_1 X_2(Q)Y_2\rangle$ remains
unchanged under the replacement $Q\to Q+\omega_r$ (the same shift for
all $Q$ in the expression) which will be true if the sum over $n$ is
convergent.

We remark that the present formalism actually only requires
$A_0(\bfx)$ to be everywhere diagonalizable but not necessarily
anti-Hermitian or even normal. This is relevant if one wants to study
the finite density case which requires introducing a chemical
potential. Effectively this amounts to add a constant Hermitian term
to $A_0$ (see e.g.\cite{GomezNicola:1994vq,Sissakian:1997cp}).

\subsection{Momentum degrees of freedom}

The momentum integration can be carried out immediately for the term
with $D_i^2$ using eq.~(\ref{eq:new.4}) and this term vanish
identically.  For the other non-gauge invariant terms of the form
$E_iD_i$, we can replace $D_i$ by $A_i$ since
$\partial_i|0\rangle=0$. Thus all remaining operators in $\langle
0||0\rangle$ are multiplicative in $\bfx$-space.

In order to integrate over $\bfk$, we can again make use of the
flavor-time basis $|\alpha,\ell\rangle$ so that only c-numbers are
involved. However, the integrand is a sum of terms of the form
$\langle XE_iX^\prime E_i\rangle$ or $\langle XE_iX^\prime
A_i\rangle$, where $X$, $X^\prime$ are functions of $Q$ and this
allows to use an equivalent and preferable method, namely, to use the
label $1$ to denote operators in the position $X$ and the label $2$
for operators in the position $X^\prime$. More generally, operators
$X^{\prime\prime}$ in the position $\langle XE_iX^\prime
E_iX^{\prime\prime}\rangle$ would also carry a label $1$ due to the
first cyclic property, eq.~(\ref{eq:cyclic.1}). That is
\begin{equation}
\langle XE_iX^\prime E_jX^{\prime\prime}\rangle=
\langle X_1X^\prime_2 X^{\prime\prime}_1 E_iE_j\rangle \,.
\end{equation}
The second cyclic property, eq.~(\ref{eq:cyclic.2}), then implies
\begin{equation}
\langle Z(Q_1,Q_2)E_iE_j\rangle = \langle
Z(Q_2,Q_1)E_jE_i\rangle\,
\label{eq:new.44a}
\end{equation}
provided that $Z(x_1,x_2)= Z(x_1+\omega_r,x_2+\omega_r)$. With this
ordering convention $Q_1$ and $Q_2$ are commuting objects and
the momentum integrals in eq.~(\ref{eq:new.43a}) can be performed
straightforwardly. All required integrals can be derived from
eq.~(\ref{eq:new.4}) and
\begin{equation}
\int\frac{d^2k}{(2\pi)^2}\frac{1}{(k^2+M_1^2)(k^2+M_2^2)} =
\frac{1}{4\pi}\frac{\log(M_1^2/M_2^2)}{M_1^2-M_2^2}
\end{equation}
A non trivial check of the calculation is that all non-gauge invariant
terms of the form $\langle XE_iX^\prime A_i\rangle$ cancel at this
step. The result can be written in the following form
\begin{equation}
\Omega_{s,2}^+(\D) = 
\frac{1}{4\pi}\sum_n\int_\Gamma
\frac{dz}{2\pi i}z^s \mu \,\tr\langle 0|
\frac{H(Q_1,Q_2)}{(Q_1-Q_2)^2}E_i^2
|0\rangle \,,
\label{eq:new.6}
\end{equation}
where
\begin{equation}
H(x_1,x_2)= \frac{1}{x_1^2-x_2^2}
\log\left(\frac{\mu^2-x_1^2}{\mu^2-x_2^2}\right)
+\frac{\mu^2-x_1x_2}{(\mu^2-x_1^2)(\mu^2-x_2^2)}\,.
\label{eq:new.71}
\end{equation}
The symmetry of $H(x_1,x_2)$ under exchange of its arguments is a
direct consequence of the cyclic property, eq.~(\ref{eq:new.44a}).

\subsection{The frequency degree of freedom}

In order to sum over frequencies, we first apply integration by parts,
eq.~(\ref{eq:39}), to transform the logarithmic term in $H(x_1,x_2)$
into a rational function. The resulting rational functions can then be
reduced to a sum of simple poles and summed over $n$ using the
identity in eq.~(\ref{eq:new.1}). The relevant formula is
\begin{eqnarray}
 \mu \sum_n && H(x_1+\omega_n,x_2+\omega_n) \simeq
\label{eq:new.72}
 \\ &&
\left(\frac{\mu}{2}
+\frac{z}{s+1}\frac{\mu}{x_1-x_2} 
+\frac{z^2}{(s+1)(s+2)}\frac{1}{x_1-x_2} 
\right)
\frac{\tanh(\mu-x_1)}
{2\mu-x_1+x_2}
+X_{1,2}+\hbox{p.p.c.}
\nonumber
\end{eqnarray}
The symbol $\simeq$ is used since this substitution only holds within
$\int_\Gamma dz z^s$ due to the integration by parts. $X_{1,2}$ means
the same expression but exchanging the labels $1$ and $2$. Finally,
p.p.c. stands for $x_{1,2}\to -x_{1,2}$. When this formula is used in
eq.~(\ref{eq:new.6}), $x_1=D_{01}$ and $x_2=D_{02}$ (where $D_{01}$
and $D_{02}$ stand for $D_0$ at positions 1 and 2 respectively) and so
p.p.c. yields the pseudo-parity conjugate.

In order to integrate over $z$ it is convenient to extract the terms
which diverge as a polynomial for large $z$:
\begin{eqnarray}
 \mu \sum_n && H(x_1+\omega_n,x_2+\omega_n) \simeq
\frac{1}{2}\Bigg[
\frac{z}{s+2}\frac{1}{x_1-x_2}
+\frac{1}{2}\frac{s+1}{s+2}
-\frac{1}{(s+1)(s+2)}\frac{m}{x_1-x_2}
\nonumber \\ &&
+\Bigg(
\frac{1}{2}\frac{s+1}{s+2}(x_1-x_2)
+\frac{s}{(s+1)(s+2)}m
+\frac{2}{(s+1)(s+2)}\frac{m^2}{x_1-x_2}
\Bigg)\frac{1}{2\mu-x_1+x_2}
\Bigg]
\nonumber \\ &&
\times \tanh(\mu-x_1)
+X_{1,2}+\hbox{p.p.c.}
\label{eq:new.8}
\end{eqnarray}

\subsection{The proper time degree of freedom}

Next we have to integrate over $z$ and apply $-d/ds|_{s=0}$. The terms
in eq.~(\ref{eq:new.8}) which are of the form $(az+b)\tanh(\omega-z)$
are immediately worked out using the function
$\Omega_\sigma(\omega,s)$. For the terms of the form
$\tanh(\omega-z)/(\omega^\prime-z)$ we will need some identities. Let
$f(z)$ be a meromorphic function such that $zf(z)\to 0$ as $z\to
\sigma\infty$, and $g(s)$ analytic at $s=0$, then
\begin{eqnarray}
\left(g(s)\int_\Gamma\frac{dz}{2\pi i}z^sf(z)\right)_{s=0}&=&0\,,
\label{eq:43}\\
-\frac{d}{ds}\left(g(s)\int_\Gamma\frac{dz}{2\pi i}z^sf(z)\right)_{s=0}&=&
g(0)\int_0^{-\sigma\infty}dt\,f(t)\,.
\label{eq:44}
\end{eqnarray}
The first equality follows from taking $s=0$ directly in the
integrand, since everything is convergent. The second equality follows
from previous one. In our case, the two terms in eq.~(\ref{eq:new.8})
which are of the form $\tanh(\omega-z)/(\omega^\prime-z)$ which are
even on the explicit $m$ are convergent for large $z$ after adding
their pseudo-parity conjugate, so eq.~(\ref{eq:44}) applies directly
to them. The remaining term is not directly convergent. It is of the
form
\begin{equation}
I_2(\omega,\omega^\prime)=-\frac{d}{ds}\left(
sg(s)\int_\Gamma\frac{dz}{2\pi i} z^s
\frac{\tanh(\omega-z)}{\omega^\prime-2z}\right)_{s=0}\,,
\end{equation}
which can be rewritten as
\begin{equation}
I_2=-\frac{d}{ds}\left(
sg(s)\int_\Gamma\frac{dz}{2\pi i}z^s
\left(\frac{\tanh(\omega-z)-\sigma}{\omega^\prime-2z}
+\frac{\sigma}{\omega^\prime-2z}\right)
\right)_{s=0}\,.
\end{equation}
The subtracted term is convergent and thus it vanishes due to
eq.~(\ref{eq:44}). The remainder can be computed explicitly
\begin{equation}
I_2=g(0)\frac{\sigma}{2}\,.
\end{equation}

The effective action takes the form
\begin{eqnarray}
W^+_2(\D) &=&
\frac{1}{8\pi}\tr\langle 0|\Bigg[
\frac{1}{2}\Omega_\sigma^\prime(m-D_{01},1)+
\left(\frac{1}{4}-\frac{1}{2}\frac{m}{D_{01}-D_{02}}\right)
	\Omega_\sigma^\prime(m-D_{01},0)
\nonumber \\ && 
+\frac{\sigma m}{4}
+ \int_0^{-\sigma\infty}dt
	\left(\frac{1}{4}(D_{01}-D_{02}) +
	\frac{m^2}{D_{01}-D_{02}}\right)
	\frac{\tanh(m-t-D_{01})}{2(m-t)-D_{01}+D_{02}}
\Bigg]
\nonumber \\ && 
\times \frac{E_i^2}{(D_{01}-D_{02})^2}|0\rangle
+X_{1,2} +\hbox{p.p.c.}
\label{eq:new.79}
\end{eqnarray}

This expression can be transformed by making some observations. First,
due to the cyclic property, the terms $X_{1,2}$ just give a factor 2
to the action. Second, recalling that $\Dh_0(X)=[D_0,X]$, the cyclic
properties imply
\begin{equation}
D_{01}-D_{02}= \Dh_{01}=-\Dh_{02}\,.
\label{eq:new.78}
\end{equation}
For instance
\begin{eqnarray}
\langle(D_{01}-D_{02})^2XY \rangle &=& 
\langle D_0^2XY-2D_0XD_0Y+XD_0^2Y \rangle \nonumber\\
&=& \langle X[D_0,[D_0,Y]]\rangle \nonumber \\
&=& \langle X\Dh_0^2Y\rangle \nonumber \\
&=& \langle \Dh_{02}^2 XY\rangle
\,.
\end{eqnarray}
Therefore, $D_{02}$ can be replaced everywhere by $D_{01}+\Dh_{02}$.
Third, because $A_0$ commutes with $\partial_0$ and
$\partial_0|0\rangle=\langle 0|\partial_0=0$, the symbol $D_{01}$ can
be replaced by $A_{01}$ everywhere. Fourth, we make use of the fields
$\A_i(x)$ which were defined in eq.~(\ref{eq:new.80}). The relation
$\Dh_0E_i=\Dh_0^2\A_i$ allows to make the replacements
\begin{equation}
E_i\frac{1}{\Dh_0^2}E_j = -\A_i\A_j\,,\quad E_i\frac{1}{\Dh_0}E_j=
-\A_i\Dh_0\A_j\,.
\end{equation}
(The minus signs come from by part integration.) Finally, we can make
use of the functions $\phi_n(\omega)$ instead of
$\Omega^\prime_\sigma(\omega,n)$ (see Appendix~\ref{app:B}). After
these manipulations, the pseudo-parity even effective action at second
order can be written as
\begin{eqnarray}
W^+_2(\D) &=&
\frac{1}{4\pi}\int dx^3\,\tr
\Bigg[
\frac{1}{2}\phi_1(m-A_0)\A_i\frac{1}{\Dh_0}\A_i
-\phi_0(m-A_0)\A_i\left(\frac{1}{4}+\frac{1}{2}\frac{m}{\Dh_0}\right)\A_i
\nonumber \\ &&
+ \int_0^{-\sigma\infty}dt
	\tanh(m-t-A_0)
	\A_i\left(\frac{1}{4}\Dh_0+\frac{m^2}{\Dh_0}\right)
	\frac{1}{2(m-t)+\Dh_0}\A_i
\Bigg] +\hbox{p.p.c.}
\label{eq:new.84}
\end{eqnarray}
(Here we are no longer using an ordering prescription with labels $1$
and $2$; the position of the operators is that given literally by the
formula.)

\subsection{Transformation properties of the result and limit cases}
\label{subsec:new.2.f}

First of all, let us show that $W^+_2(\D)$ does not contribute
to the gauge or parity anomalies. Because the equation defining $\A_i$
(eq.~(\ref{eq:new.80})) is gauge and parity covariant, we can choose
$\A_i$ to transform covariantly too. The gauge invariance under
time-independent transformations is immediate since $A_0$, $\A_i$ and
$\Dh_0$ transform covariantly.  Under discrete transformations, $\A_i$
and $\Dh_0$ remain invariant, whereas $A_0$ changes by integer
multiples of $i\pi$ thus the integral term in $W^+_2(\D)$ is invariant
due to periodicity of the hyperbolic tangent. For the terms containing
$\phi_n$ the invariance under discrete gauge transformations follows
from eqs.~(\ref{eq:new.54}) through a cancellation among the terms
with $\phi_0$, $\phi_1$ and their p.p.c. Since the cancellation is non
trivial it provides a check of the calculation.

The parity invariance of the terms containing $\phi_n$ in $W^+_2(\D)$
follows immediately from the parity properties of these functions,
eq.~(\ref{eq:new.52.c}). On the other hand, in the term containing the
integral over $t$, a parity transformation can be seen to be
equivalent to change $\sigma\to -\sigma$, in agreement with
eq.~(\ref{eq:18b}). Thus the parity violating contribution would come
from extending the integral on $t$ to the range $-\infty < t <
\infty$. Taking into account the p.p.c. term, such integral vanishes
because it is convergent and the integrand is an odd function of
$m-t$.

Next, let us check that $W^+_2(\D)$ only depends on $E_i(x)$ and not on
the particular solution taken for the field $\A_i(x)$. Recalling the
definition of $\A_i$ (cf. eq.~(\ref{eq:new.80})) this requires that
$\A_i\A_i$ should appear with at least two powers of the operator
$\Dh_0$. The identity $\Dh_{02}=Q_2-Q_1$ (eq.~(\ref{eq:new.78})) then
requires in eq.~(\ref{eq:new.6}) that the function $H(x_1,x_2)$
should be of order $(x_1-x_2)^2$ as $x_1-x_2\to 0$. In fact this is
the case,
\begin{equation}
\lim_{x_1,x_2\to x}\frac{H(x_1,x_2)}{(x_1-x_2)^2} = 
\frac{3\mu^2+x^2}{6(\mu^2-x^2)^3} \,.
\label{eq:new.85}
\end{equation}
In eq.~(\ref{eq:new.84}) the dependence through $E_i$ is not manifest
because we have used the symmetry between the labels $1$ and $2$ to
simplify the expression. To check this property directly from the
final expression for $W^+_2(\D)$ it is necessary to restore $D_{01}$,
$D_{02}$, add $X_{1,2}$ (dividing by two), eliminate
$D_{02}$ in favor of $\Dh_0$ and then expand in powers
of $\Dh_0$. (This procedure is carried out in detail for $W^-_2$ in
section~\ref{sec:4}.) After doing this, it is indeed found that the
necessary cancellations take place and only terms of order
$\A_i\Dh_0^2\A_i$ survive. These cancellations among different terms
are again a non trivial check of the result. Retaining the lowest
order in the expansion on $\Dh_0$ just described yields
\begin{equation}
W^+_{2,\,{\rm leading}}(\D)=
-\frac{1}{24\pi}\int d^3x\,\tr\,\Big[
\frac{\tanh(m-A_0)}{2m}
-\frac{1}{4\cosh^2(m-A_0)}
\Big]E_i^2 + \hbox{p.p.c.}
\label{eq:new.86}
\end{equation}
It is noteworthy that the terms in the expansion combine in such a way
that the integral over $t$ present in the original expression can be
done in closed form by integration by parts. It has been checked that
the same final expression is obtained starting directly from
eq.~(\ref{eq:new.85}).

In fact, the interest of $W^+_{2,\,{\rm leading}}$ goes beyond
checking explicitly the independence of $W^+_2$ on the spurious
degrees of freedom contained in the variables $\A_i$.  $W^+_{2,\,{\rm
leading}}$ has been obtained by making first an expansion in powers of
$D_i$ to second order and then keeping the leading order in an
expansion in powers of $\Dh_0$. As noted at the end of
subsection~\ref{subsec:2.g}, this double expansion in $D_i$ and
$\Dh_0$ is the natural finite temperature version of the standard
space-time gradient expansion at zero temperature (which in the
absence of other non-gauge fields, is also equivalent to a $1/m$
expansion). No symmetry has been broken by the further expansion in
powers of $\Dh_0$ and $W^+_{2,\,{\rm leading}}$ is manifestly gauge
and parity invariant. On the other hand, expanding in powers of $A_0$,
as implied by standard perturbation theory, would break invariance
under discrete gauge transformations.

In the particular case of Abelian and stationary fields (i.e.,
time-independent $\bfA$) $\Dh_0$ vanishes so $W^+_{2,\,{\rm leading}}$
does in fact coincide with $W^+_2$.

It is also of interest to examine the zero temperature limit of
$W^+_2(\D)$. As shown in 
Appendix~\ref{app:B},
\begin{equation}
\lim_{\beta\to\infty}\phi_{n-1}(\omega) =
\frac{1}{n!}\omega^{n}\varepsilon(m)\,,\quad n=0,1,2,\dots \,.
\end{equation}
(Recall that we are using units $\beta=2$ so the zero temperature
limit corresponds to large $\omega=m-A_0$). After some non trivial
cancellations, one finds 
\begin{equation}
W^+_2(\D)=
-\frac{1}{4\pi}\int d^3x\,\tr\,
E_i\left[\frac{1}{8|m|}+\frac{1}{2}(4m^2+\Dh^2_0)\int_{|m|}^{+\infty}dt
\frac{1}{4t^2(4t^2-\Dh_0^2)}\right]E_i\,, \quad (T=0) \,.
\end{equation}
This result depends only on $E_i$ (and not $\A_i$) and is free of
parity or gauge anomalies. Full Lorentz invariance, which is recovered
at zero temperature, is no supported since it has been broken by
making an expansion on the spatial gradient only. In order to compare
with direct zero temperature calculations, we will only retain terms
of dimension 4, or equivalently, the leading term in a $1/m$
expansion. This gives
\begin{equation}
W^+_2(\D)=
-\frac{1}{24\pi}\frac{\varepsilon(m)}{m}\int d^3x\,\tr\,
E_i^2 + O\left(\frac{1}{m^3}\right)\,, \quad (T=0) \,.
\end{equation}
(Of course, the same result is obtained taking the zero temperature
limit of $W^+_{2,\,{\rm leading}}(\D)$ in eq.~(\ref{eq:new.86})). Then
Lorentz invariance implies
\begin{equation}
W^+(\D)=
-\frac{1}{48\pi}\frac{\varepsilon(m)}{m}\int d^3x\,\tr\,
F_{\mu\nu}^2 + O\left(\frac{1}{m^3}\right)\,, \quad (T=0) \,.
\end{equation}
(Up to constant terms coming from $W_0(\D)$.) It is an interesting
check of the calculation that precisely the same result is obtained
from a direct zero temperature calculation (using for instance the
$1/m$ expansion coefficients given in \cite{Salcedo:1996qy}).

\section{The pseudo-parity odd effective action at second order}
\label{sec:4}

In this section we will compute the leading contribution to the
imaginary part of the effective action, which comes from the second
order in the gradient expansion. This leading term is particularly
interesting since it contains the essential parity anomaly as well as
the multivaluation under large gauge transformations.

Many of the ideas needed to carry out the calculation have already
appeared in the previous section, so they do not need to be repeated
here. We start from eq.~(\ref{eq:new.57}) (recall that we use units
$\beta=2$). Use of eq.~(\ref{eq:new.3}) allows to compute the Dirac
trace. Keeping just the terms with $\epsilon_{ij}$ yields
\begin{eqnarray}
\Omega_{s,2}^-(\D) &=& -i\eta\int\frac{d^2k}{(2\pi)^2}\sum_n\int_\Gamma
\frac{dz}{2\pi i}z^s\epsilon_{ij}\tr\langle 0|\left(
	\frac{Q}{\Delta}D_i\frac{Q}{\Delta}D_j\frac{Q}{\Delta}
\right.
\nonumber\\
&&
\left.
	+(\mu^2-k^2) \frac{1}{\Delta}D_i\frac{Q}{\Delta}D_j\frac{1}{\Delta}
	-\mu^2 \frac{1}{\Delta}D_i\frac{1}{\Delta}D_j\frac{Q}{\Delta}
	-\mu^2 \frac{Q}{\Delta}D_i\frac{1}{\Delta}D_j\frac{1}{\Delta}
\right)|0\rangle
\,,
\end{eqnarray}
where the trace refers to flavor only. (This expression is analogous
to that in eq.~(\ref{eq:new.58}) for the pseudo-parity even part.)

Next, the spatial covariant derivatives are brought to the right. It
has been found to be best to move $D_i$ first and then $D_j$ to
generate a smaller number of terms. This involves the quantities $E_i$
and $F_{ij}=[D_i,D_j]$, and also $A_i$ appears due to
$D_i|0\rangle=A_i|0\rangle$. In this form the following expression is
derived:
\begin{eqnarray}
\Omega_{s,2}^-(\D) &=& 
-i\eta\int\frac{d^2k}{(2\pi)^2}\sum_n\int_\Gamma
\frac{dz}{2\pi i}z^s\tr\langle 0|\left(
-(\mu^2+k^2)\frac{Q}{\Delta^3}F+\frac{Q^3}{\Delta^3}F
\right. \nonumber\\ &&
-\frac{1}{\Delta}E\frac{Q}{\Delta^2}E
+\mu^2\frac{1}{\Delta}E\frac{Q}{\Delta^3}E
-\frac{1}{\Delta}E\frac{Q^3}{\Delta^3}E
-\mu^2\frac{1}{\Delta^3}E\frac{Q}{\Delta}E
-\frac{Q}{\Delta}E\frac{Q^2}{\Delta^3}E
\nonumber\\ &&
+(\mu^2+2k^2)\frac{Q}{\Delta^2}E\frac{Q^2}{\Delta^3}E
-(\mu^2+2k^2)\frac{Q}{\Delta^3}E\frac{Q^2}{\Delta^2}E
-\mu^2\frac{1}{\Delta^3}E\frac{Q^3}{\Delta^2}E
\nonumber\\ &&
-\frac{Q}{\Delta^2}E\frac{Q^4}{\Delta^3}E
-2\frac{Q^2}{\Delta^2}E\frac{Q^3}{\Delta^3}E
+\frac{Q^2}{\Delta^3}E\frac{Q^3}{\Delta^2}E
+k^2\frac{1}{\Delta^2}E\frac{Q^3}{\Delta^3}E
\nonumber\\ &&
-(\mu^2-k^2)\frac{1}{\Delta}E\frac{1}{\Delta^2}A
+(\mu^2+k^2)\frac{Q}{\Delta^2}E\frac{Q}{\Delta^2}A
+\mu^2\frac{1}{\Delta^2}E\frac{Q^2}{\Delta^2}A
\nonumber\\ & & \left.
+k^2\frac{Q^2}{\Delta^2}E\frac{1}{\Delta^2}A
-\frac{Q}{\Delta}E\frac{Q}{\Delta^2}A
-\frac{Q^2}{\Delta^2}E\frac{Q^2}{\Delta^2}A
-\frac{Q^3}{\Delta^2}E\frac{Q}{\Delta^2}A
\right)|0\rangle
\,.
\label{eq:43a}
\end{eqnarray}
This expression is analogous to that in eq.~(\ref{eq:new.43a}) for
$\Omega_{s,2}^+(\D)$ and many of the remarks made there apply here
too.  A standard differential form notation has been adopted for the
spatial indices, i.e., $dx_1$ and $dx_2$ are anticommuting,
$d^2x\epsilon_{ij}=dx_idx_j$, $D=D_idx_i$, $A=A_idx_i$, $E=E_idx_i$
and $F=D^2$. The operator inside $\langle 0||0\rangle$ is
multiplicative in $\bfx$-space, thus the part of $\langle 0||0\rangle$
corresponding to that space has been replaced by $\int d^2x$ and
$d^2x$ has been included in the differential forms, so $|0\rangle$ now
refers to the $x_0$-space only. The terms with explicit $A$ break
gauge invariance and will cancel later. Finally, to obtain a more
compact expression, the two cyclic properties have been used.

To carry out the momentum integration, we again make use of the method of
labelling the $Q$ operators with labels $1$ or $2$ according to their
position. Note, however, that now
\begin{equation}
\langle Z(Q_1,Q_2)E^2\rangle =-\langle Z(Q_2,Q_1)E^2\rangle\,,
\label{eq:44a}
\end{equation}
the extra minus sign coming from $E$ being a 1-form. Since all
quantities are now commuting (or anti-commuting) momentum integration
is straightforward.  A non trivial check of the calculation is that all
non-gauge invariant terms $\langle X_1EX_2A\rangle$ cancel at this
step. The result can be written in the following form
\begin{equation}
\Omega_{s,2}^-(\D) = 
\frac{i\eta}{4\pi}\sum_n\int_\Gamma
\frac{dz}{2\pi i}z^s\tr\langle 0|\left(
\frac{Q}{\mu^2-Q^2}F-
\frac{1}{2}\frac{H(Q_1,Q_2)}{Q_1-Q_2}E^2
\right)|0\rangle \,,
\label{eq:new.95}
\end{equation}
where $H(x_1,x_2)$ is the same function as in the pseudo-parity even
case\footnote{This suggests that the second order pseudo-parity even
and odd components are closely related and perhaps connected through
some kind of analytical extrapolation. Such relations exist in 1+1
dimensions and allow to exactly compute the corresponding Weyl
determinant\cite{Rothe:1986br}.}, eq.~(\ref{eq:new.71}). Note that in
this case $E^2$ is odd under exchange of the labels $1$ and $2$ (since
$E$ is a 1-form), so again the total expression is even under this
exchange.

In order to sum over frequencies we need a new identity for the term
with $F$, namely\footnote{The series in the left-hand side is not
absolutely convergent, nevertheless it will be so upon subtraction of
a suitable $z$-independent term, as in eq.~(\ref{eq:new.30}). Thus
this equality holds inside $\int_\Gamma dz z^s$.}
\begin{equation}
\sum_n\frac{x+\omega_n}{\mu^2-(x+\omega_n)^2} \simeq
\frac{1}{2}\tanh(\mu-x)-\hbox{p.p.c.}\,.
\end{equation}
In the term with $H(Q_1,Q_2)$ we again use integration by parts to
eliminate the logarithm. This yields
\begin{equation}
\sum_n H(x_1+\omega_n,x_2+\omega_n) \simeq
\left(\frac{z}{s+1}\frac{1}{x_1-x_2} + \frac{1}{2}\right)
\frac{\tanh(\mu-x_1)}{2\mu-x_1+x_2} + X_{1,2}+ \hbox{p.p.c.}\,.
\end{equation}
Note that this result multiplied by $\mu=m-z$ needs to coincide with
that in eq.~(\ref{eq:new.72}) only inside $\int_\Gamma dz z^s$, since
integration by parts has been used in both cases. In analogy to 
eq.~(\ref{eq:new.8}), it is convenient to rewrite this expression as
\begin{eqnarray}
\sum_n H(x_1+\omega_n, && x_2+\omega_n)  \simeq 
\nonumber\\ &&
\Bigg[
-\frac{1}{2}\frac{1}{s+1}\frac{1}{x_1-x_2}
+\Bigg(
\frac{1}{2}\frac{s}{s+1}
+\frac{1}{s+1}\frac{m}{x_1-x_2}
\Bigg)\frac{1}{2\mu-x_1+x_2}
\Bigg] \nonumber \\
&& \times \tanh(\mu-x_1) +X_{1,2}+\hbox{p.p.c.}
\end{eqnarray}

After carrying out the $z$-integration, we arrive at a formula
analogous to eq.~(\ref{eq:new.79}), namely

\begin{eqnarray}
W^-_2(\D) &=&
\frac{i\eta}{4\pi}\tr\langle 0|\Bigg[
\frac{1}{2}\Omega^\prime_\sigma(m-D_0,0)F
+\frac{1}{2}
\Bigg(\frac{1}{2}\frac{1}{\Dh_{02}}\Omega^\prime_\sigma(m-D_{01},0)
\nonumber \\ && 
+\frac{\sigma}{4}
-\frac{m}{\Dh_{02}}\int_0^{-\sigma\infty}dt
	\frac{\tanh(m-t-D_{01})}{2(m-t)+\Dh_{02}} 
+X_{1,2}\Bigg) \frac{E^2}{\Dh_{02}}
\Bigg]|0\rangle
 -\hbox{p.p.c.} 
\label{eq:new.99}
\end{eqnarray}
The contribution $X_{1,2}$ refers to everything inside the parenthesis
and it gives a factor 2 due to the cyclic property. The final form is
obtained by further replacing $D_{01}$ by $A_{01}$, $E^2$ by
$-\Dh^2_{02}\A^2$ and $\langle 0|0\rangle$ by $\int dx_0$. This gives
\begin{eqnarray}
W_2^-(\D) &=&
\frac{i\eta}{8\pi}\int dx_0\tr\Bigg[
(F-\A^2)\Phi_\sigma
-\sigma\A\Dh_0\A \nonumber\\
&&
+2m\int_0^{-\sigma\infty} dt \A\left(
\frac{\tanh(m-t+A_0)}{2(m-t)+\Dh_0} -
\frac{\tanh(m-t-A_0)}{2(m-t)-\Dh_0}
\right)\A
\Bigg],
\label{eq:8}
\end{eqnarray}
where $\Phi_\sigma=\Omega_\sigma^\prime(m-A_0,0)-{\rm p.p.c.}$ was
introduced in eq.~(\ref{eq:new.39}). Note that $d^2x$ is included in
the differential forms. 

In order to obtain the leading term in a further expansion in powers
of $\Dh_0$ (the analogous of $W^+_{2,\,{\rm leading}}(\D)$) we go back
to the symmetrized formula, eq.~(\ref{eq:new.99}), then express the
operator $D_{02}$ in $X_{1,2}$ as $D_{01}+\Dh_{02}$ and carry out an
expansion in powers of $\Dh_{02}$ keeping the leading order.  At the
end $D_{01}$ can be replaced by $A_{01}$. This needs only be done in
the term containing the integral over $t$:
\begin{eqnarray}
&& 
m\int_0^{-\sigma\infty} dt \left[
\frac{\tanh(m-t-D_{01})}{2(m-t)+\Dh_{02}} -X_{1,2}
\right]
 \nonumber\\
&& 
m\int_0^{-\sigma\infty} dt \left[ 
\frac{\tanh(m-t-D_{01})}{2(m-t)+\Dh_{02}}
-
\frac{\tanh(m-t-D_{01}-\Dh_{02})}{2(m-t)-\Dh_{02}}
\right]
 \nonumber\\
&=&
m\int_0^{-\sigma\infty} dt \left[ 
-\frac{\tanh(m-t-D_{01})}{2(m-t)^2}
+\frac{1}{\cosh^2(m-t-D_{01})}\frac{1}{2(m-t)}
\right]\Dh_{02} + O(\Dh_{02}^2)
 \nonumber\\
&=&
\frac{1}{2}\tanh(m-D_{01})\Dh_{02} + O(\Dh_{02}^2)\,.
\end{eqnarray}
So finally the result is
\begin{eqnarray}
W_{\rm anom}(\D) &=& \frac{i\eta}{8\pi}\int dx_0 \tr\Bigg[
(F-\A^2)\Phi_\sigma
\nonumber\\ && \quad
+\frac{1}{2}\Big(\tanh(m-A_0)+
\tanh(m+A_0)-2\sigma\Big)\A\Dh_0\A \Bigg]\,.
\label{eq:22}
\end{eqnarray}

Let us finish by discussing the properties of $W^-_2$ paralleling the
discussion of $W_{\rm anom}$ in section~\ref{sec:summary}. Using the
same arguments as for $W_{\rm anom}$ it follows that $W^-_2$ is well
defined, i.e., it does not depend on the choice of $\A_i(x)$. This
implies that it is possible to express all terms not contained in
$W_{\rm anom}$ in terms of the electric field.

The term of $W^-_2$ with the integral over $t$ is actually independent
of $\sigma$ since using $+\infty$ or $-\infty$ as the upper limit of
the integral gives the same result. This follows because the integrand
is an odd function of $m-t$. Therefore it is parity invariant and
$W^-_2$ has the same parity anomaly as $W_{\rm anom}$.
Likewise, the difference between $W^-_2$ and $W_{\rm anom}$ is
strictly gauge invariant.

Eq.~(\ref{eq:new.21}) for Abelian-like configurations also holds for
$W^-_2$, since the terms without $F$ cancel. In the massless case, the
term with an integral over $t$ cancels and so eq.~(\ref{eq:new.23})
still holds for $W^-_2$.

The zero temperature limit of $W^-_2$ is given by
\begin{equation}
W_2^-(\D) =\eta\sigma\Theta(-\sigma m)W_{\rm CS}
-\frac{i\eta}{8\pi}\int dx_0\tr\int_{|m|}^{\infty}
dt E\frac{m\Dh_0}{t^2(4t^2-\Dh_0^2)}E\,,
\quad (T=0)\,.
\label{eq:18}
\end{equation}
The first term was already present in $W_{\rm anom}$. Since the second
term is of order $1/m^2$ in an inverse mass expansion, it is
ultraviolet finite.

The zero temperature limit of $W_2^-(\D)$ is not directly Lorentz
invariant. Restoring this invariance requires including higher order
terms in the gradient expansion. Nevertheless, keeping the first non
trivial $1/m$ contribution in eq.~(\ref{eq:18}), the only Lorentz
invariant completion is
\begin{equation}
W^-(\D)= \eta\sigma\Theta(-\sigma m)W_{\rm CS}
-\frac{i\eta}{8\pi}\frac{\varepsilon(m)}{12m^2}\int d^3x
\epsilon_{\mu\nu\alpha}\tr(F_{\beta\mu}\Dh_\alpha F_{\beta\nu})
+O\left(\frac{1}{m^3}\right) \,, \quad (T=0) \,.
\end{equation}
Precisely the same result is obtained through a direct zero
temperature calculation using the formalism of
Ref.\cite{Salcedo:1996qy}. This is another non trivial test of the
calculation.

\section*{Acknowledgments}
I gratefully acknowledge a critical reading of the manuscript by
C. Garc\'{\i}a-Recio. This work is supported in part by funds provided
by the Spanish DGICYT grant no. PB95-1204 and Junta de Andaluc\'{\i}a
grant no. FQM0225.

\appendix

\section{The $A_0$-stationary gauges}
\label{app:A}

Let the space-time have the topology $M_{d+1}=M_d\times{\rm S}^1$, $d$
being the space dimension. All functions involved are assumed to be
continuous on $M_{d+1}$ (and thus, in particular, periodic in the time
coordinate) unless otherwise stated. Let $A_\mu(x)$ be a gauge
configuration on $M_{d+1}$. We want to show that there exists another
configuration $B_\mu(x)$ in the $A_0$-stationary gauge (i.e.,
$\partial_0B_0=0$) related to $A_\mu$ by a gauge transformation. To do
this, let us construct the following time-independent field
\begin{equation}
\Omega(\bfx)= T\exp\left(-\int_0^\beta A_0({\bfx},x_0)dx_0\right)\,.
\end{equation}
(The symbol $T$ stands for time ordered product.) As is well-known, $\Omega$ transforms covariantly (in fact $\tr(\Omega)$
is just the Polyakov loop)
\begin{equation}
\Omega^U(\bfx)= U^{-1}(\bfx,0)\Omega(\bfx)U(\bfx,0)\,.
\end{equation}
Thus, $\Omega$ is gauge invariant modulo stationary gauge
transformations. In view of this, we demand that $B_\mu$ should yield
the same field $\Omega$ as $A_\mu$, that is,
\begin{equation}
\Omega(\bfx)= \exp\left(-\beta B_0({\bfx}\right))\,.
\end{equation}
That the function $B_0(\bfx)$ above exists (with continuity on $M_d$)
follows from $\Omega$ being an homotopically trivial application from
$M_d$ into $G$. In fact,
\begin{equation}
U_1(\bfx,t)= T\exp\left(-\int_0^tA_0({\bfx},t^\prime)dt^\prime\right)\,,\quad 0\le
t\le\beta 
\end{equation}
defines one such deformation of $\Omega$ to the identity element of
$G$ with $t$ as interpolating parameter. It only remains to show that
there actually exists a gauge transformation bringing $A_\mu$ to
$B_\mu$. This is achieved with
\begin{equation}
U(\bfx,x_0) =U_1(\bfx,x_0)\exp\left(x_0B_0(\bfx)\right)\,.
\end{equation}
The first factor, $U_1$, brings $A_0$ to zero but breaks periodicity
in Euclidean time. The second factor $e^{x_0B_0}$ yields $A^U_0=B_0$
and reestablishes the continuity of $U$ on $M_{d+1}$. Therefore, a
$A_0$-stationary gauge always exists.

Next, let us determine the remaining gauge freedom within an
$A_0$-stationary gauge. Let $A_\mu$ and $B_\mu$ be two
$A_0$-stationary configurations and let $U$ be a gauge transformation
bringing $A_\mu$ to $B_\mu$. Then
\begin{equation}
B_0(\bfx)=U^{-1}(x)\partial_0U(x)+U^{-1}(x)A_0(\bfx)U(x)\,.
\end{equation}
In terms of the auxiliary variable $V(x)=\exp(x_0A_0(\bfx))U(x)$, the previous
equation becomes
\begin{equation}
B_0(\bfx)=V^{-1}(x)\partial_0V(x)\,.
\end{equation}
Its general solution is
\begin{equation}
V(x)=U_0(\bfx)\exp(x_0B_0(\bfx))\,,
\end{equation}
where $U_0$ is an arbitrary time-independent gauge transformation. In
terms of $U(x)$
\begin{equation}
U(x)= \exp(-x_0A_0(\bfx))U_0(\bfx)\exp(x_0B_0(\bfx))\,.
\end{equation}
This is the most general ``gauge transformation'' bringing $A_\mu$
into a $A_0$-stationary gauge, and depends on the two arbitrary time-independent fields $B_0$
and $U_0$. However, $U$ will not be a true gauge transformation unless
it is also continuous on $M_{d+1}$. To impose this restriction, let us change
variables from $B_0(\bfx)$ to $\Lambda(\bfx)$ by means of
\begin{equation}
B_0(\bfx)=U_0^{-1}(\bfx)(A_0(\bfx)+\Lambda(\bfx))U_0(\bfx)\,.
\end{equation}
This allows to write $U(x)$ more conveniently as
\begin{equation}
U(x)= \exp(-x_0A_0(\bfx))\exp(x_0(A_0(\bfx)+\Lambda(\bfx)))U_0(\bfx)\,.
\end{equation}
The continuity of $U(x)$ follows from requiring
$U({\bfx},0)=U(\bfx,\beta)$, which is equivalent to
\begin{equation}
\exp(\beta A_0(\bfx))=\exp(\beta (A_0(\bfx)+\Lambda(\bfx)))\,.
\end{equation}

Finally, let us show that for simply connected gauge groups, it is not
always possible to bring any given gauge field configuration to the
$A_0$-stationary gauge using only small transformations. To show this,
let us assume that the initial configuration $A_\mu$ is already
$A_0$-stationary. Let us perform a large gauge transformation on it,
yielding the configuration $A^{\prime\prime}_\mu$, and then bring this
configuration again to an $A_0$-stationary gauge to finally obtain a
configuration $A^\prime_\mu$. Without loss of generality it can be
assumed that $A_\mu$ and $A^\prime_\mu$ are related by a discrete
transformation of $A_0(\bfx)$ (i.e., with no time-independent part).
We want to show that the gauge transformation bringing
$A^{\prime\prime}_\mu$ to $A^\prime_\mu$ cannot always be chosen to be
small. Indeed, if it were small, $A_\mu$ and $A^\prime_\mu$ would be
related by a large gauge transformation which is also a discrete
transformation for $A_0(\bfx)$. However, this not always possible if
the gauge group is simply connected. This can be seen by taking
$A_0(\bfx)$ to be a constant diagonal matrix such that the spectrum of
$\exp(\beta A_0(\bfx))$ is not degenerated. Then, the allowed
$\Lambda(\bfx)$ for this configuration will be also diagonal and
constant and thus $e^{x_0\Lambda}$ will necessarily be small, that is,
no large discrete transformation exists for such $A_0(\bfx)$. On the
other hand, in the Abelian U(1) case, it is always possible to deform
$A_0(x)$ to a time-independent one preserving the Polyakov loop and no
large transformations are required.

\section{Auxiliary functions}
\label{app:B}

The functions $\Omega_\Gamma(\omega,s)$ are defined through the formula
\begin{equation}
\Omega_\Gamma(\omega,s) = -\int_\Gamma \frac{dz}{2\pi i}z^s
\tanh(\omega-z) \,.
\label{eq:new.B1}
\end{equation}
The integral is defined for ${\rm Re}(s)< -1$ and extends to a
meromorphic function of $s$ with a simple pole at $s=-1$.  The
integration path $\Gamma$ follows the ray $\theta$ starting from
infinity, encircles zero clockwise and goes back to infinity along the
ray $\theta-2\pi$ on the $z$-complex plane.  Because the function
$\tanh(z)$ has simple poles at $z=\omega_n= i\pi(n+\frac{1}{2})$, the
integral is not defined when $\omega+\omega_n$ lies on $\Gamma$ for
some $n$. As a consequence, the complex plane corresponding to the
variable $\omega$ is cut along rays with angle $\theta$ stemming from
the points $\omega=\omega_n$ which are branching points of the
function $\Omega_\Gamma(\omega,s)$ (unless $s$ is an integer).  Let us
denote the $\omega$-complex plane so cut by ${\rm C}_\theta$, and the
corresponding Riemann surface by $\Ct$. This latter manifold is common
to all values of $\theta$.

The angle $\theta$ can take any real value excepting $\theta
=\pm\pi/2$~(mod. $2\pi$) (i.e., $\Gamma$ on the imaginary axis). The
function $z^s$ is defined taking $\arg(z)$ in the range
$]\theta-2\pi,\theta[$. Correspondingly, we will use the notation
$(z)^s_\theta$, $\arg_\theta(z)$ and
$\Omega_\theta(\omega,s)$. Applying Cauchy's theorem, the
function can then be written as
\begin{equation}
\Omega_\theta(\omega,s) = \sum_{n\in{\rm Z}}(\omega-\omega_n)^s_\theta
\,.
\label{eq:new.B2}
\end{equation}
If $\theta$ is shifted to $\theta^\prime$ but without crossing the
limits $\pm\pi/2$ (mod. $2\pi$), only a finite number of terms in the
series get modified and the two functions $\Omega_\theta$,
$\Omega_{\theta^\prime}$ are related by analytical continuation, that
is, they coincide on the $\Ct$. On the other hand, from the definition
it follows immediately that
\begin{equation}
\Omega_{\theta+2\pi n}(\omega,s) =
e^{2\pi ins}\Omega_{\theta}(\omega,s)\,, \quad (n\in\hbox{Z})
\label{eq:new.B3}
\end{equation}
As will be shown below, $\Omega_\Gamma(\omega,k)=0$ for
$k=0,1,2,\dots$, so $\Omega^\prime_{\theta+2\pi n}(\omega,k)=
\Omega^\prime_\theta(\omega,k)$ (where the prime refers to the
derivation on $s$). Thus, all angles differing by an integer multiple
of $2\pi$ give the same effective action and are equivalent. It
follows that all practical cases are covered by taking $\theta=\pi$
and $\theta=2\pi$, which correspond to $\sigma=1$ and $\sigma=-1$
respectively.

The $\omega$-complex planes $\C_\sigma$ are cut along rays which are
parallel to the real axis, start at $\omega=\omega_n$ and go to the
right for $\sigma=+1$ and to the left for $\sigma=-1$. To describe the
functions $\Omega_\sigma(\omega,s)$ on the Riemann surface $\Ct$, it
will prove convenient to introduce the $\omega$-complex plane cut in
another way, which will be denoted by $\C_p$ and is defined as
follows. Starting from any of $\C_\sigma$ planes, rotate upwards the
rays which are above the real axis so that these rays lie now on the
positive imaginary axis, and likewise rotate downwards the rays below
the real axis to put them on the negative imaginary axis. Since the
branching points which are nearest to the origin are
$\omega=\pm\omega_0=\pm i\pi/2$, the functions
$\Omega_\sigma(\omega,s)$ extended to $\C_p$ are analytical for all
$\omega$ except when $\omega$ is purely imaginary and above $\omega_0$
or below $-\omega_0$. The plane $\C_p$ has the property of supporting
parity transformations, which correspond to $\omega\to -\omega$ (since
in practice $\omega= m\pm A_0$) and also discrete gauge
transformations, $\omega\to\omega+i\pi$, where the two points are
connected by a straight line. Besides, for each value of $\sigma$, the
function $\Omega_\sigma(\omega,s)$ takes the same value on $\C_\sigma$
and $\C_p$ when $\omega$ is in the half-plane $\sigma{\rm
Re}(\omega)>0$.

The functions $\Omega_\sigma(\omega,s)$ can be directly related to the
Hurwitz function $\zeta(z,q)$ \cite{Gradshteyn:1980}~(p. 1073 and
ff.):
\begin{equation}
\zeta(z,q)= \sum_{n=0}^\infty (q+n)^{-z}_\pi \,,
\end{equation}
where the subindex $\pi$ means $|\arg(q+n)|<\pi$. This function is
analytical (as a function of $q$) except on the negative real
axis. The series representation of $\Omega_\sigma$,
eq.~(\ref{eq:new.B2}), allows to write
\begin{eqnarray}
\Omega_+(\omega,s) &=& 
(-i\pi)^s\zeta\left(-s,\frac{1}{2}-\frac{\omega}{i\pi}\right) +
(i\pi)^s\zeta\left(-s,\frac{1}{2}+\frac{\omega}{i\pi}\right)
 \nonumber \\
\Omega_-(\omega,s) &=& 
e^{2\pi is}(-i\pi)^s\zeta\left(-s,\frac{1}{2}-\frac{\omega}{i\pi}\right) +
(i\pi)^s\zeta\left(-s,\frac{1}{2}+\frac{\omega}{i\pi}\right)\,,
\label{eq:new.B4}
\end{eqnarray}
where all arguments are to be taken on $]-\pi,\pi[$ and the equalities
hold on $\C_p$. An immediate consequence is that the functions
$\Omega_+(\omega,s)$ and $\Omega_-(\omega,s)$ coincide everywhere when
$s$ is an integer. Using the identities\cite{Gradshteyn:1980}
\begin{equation}
\zeta(-n,q)=-\frac{B_{n+1}(q)}{n+1}\,,\quad
B_n(1-x)=(-1)^nB_n(x)\,,\quad n=0,1,2,\dots
\end{equation}
where $B_n(x)$ are the Bernoulli polynomials, it follows immediately
from eq.~(\ref{eq:new.B4}) that
\begin{equation}
\Omega_\pm(\omega,n)=0\,,\quad n=0,1,2,\dots
\label{eq:new.B7}
\end{equation}
and furthermore,
\begin{eqnarray}
\Omega^\prime_+(0,n)&=&\Omega^\prime_-(0,n)\,,\quad n=0,2,4,\dots 
\label{eq:new.B8a}\\
\Omega^\prime_\sigma(0,n)&=&
-\sigma(i\pi)^{n+1}\frac{B_{n+1}(\frac{1}{2})}{n+1}\,,\quad
n=1,3,5,\dots
\label{eq:new.B8b}
\end{eqnarray} 
where the prime refers to derivative with respect to $s$. Another
useful property follows from the identity
\begin{equation}
\partial_\omega\Omega^\prime_\sigma(\omega,s)=
\Omega_\sigma(\omega,s-1)+s\Omega^\prime_\sigma(\omega,s-1) \,,
\end{equation}
which implies
\begin{equation}
\partial_\omega\Omega^\prime_\sigma(\omega,n+1)=
(n+1)\Omega^\prime_\sigma(\omega,n) \,,\quad n=0,1,2,\dots
\end{equation}

The functions  $\Omega_\sigma^\prime(\omega,n)$ for non negative
integer $n$ can be expressed in terms of simple integrals  as follows
\begin{eqnarray}
\Omega_\sigma^\prime(\omega,n)&=&-\frac{d}{ds}\left(
\int_\Gamma\frac{dz}{2\pi i}z^s\tanh(\omega-z)\right)_{s=n}
\nonumber \\ &=&
-\frac{d}{ds}\left(
\int_\Gamma\frac{dz}{2\pi i}z^s
\left[z^n(\tanh(\omega-z)-\sigma)\right]\right)_{s=0}
\nonumber\\
&=&\int_0^{-\sigma\infty}dt\,t^n(\tanh(\omega-t)-\sigma)
\nonumber\\
&=&\int_{\sigma\infty}^\omega dt\,(\omega-t)^n(\tanh(t)-\sigma) \,,
\quad n=0,1,2,\dots
\label{eq:new.B12}
\end{eqnarray}
The defining integral has been converted into a convergent one by
using eq.~(\ref{eq:42}) and then eq.~(\ref{eq:44}) has been applied.
This result refers to $\omega$ in $\C_\sigma$ and in the integrals
over $t$, the integration path is to be taken parallel to the real
axis. (A similar method would also yield the relations in
eqs.~(\ref{eq:new.B7},\ref{eq:new.B8a},\ref{eq:new.B8b}).)  In particular, for
$n=0$ the integral can be done in closed form
\begin{eqnarray}
\Omega_\sigma^\prime(\omega,0) &=& \log(e^{-2\sigma\omega}+1)\,,
\end{eqnarray}
which can be rewritten as
\begin{eqnarray}
\Omega^\prime_\sigma(\omega,0) &=& \phi_0(\omega)-\sigma\omega\,, 
\label{eq:new.B9} \\
\phi_0(\omega)&=& \log(2\cosh(\omega)) \,.
\end{eqnarray}

Next, let us introduce the functions $\phi_n(\omega)$ for arbitrary
integer $n$. They are defined on $\C_p$ by the relations
\begin{eqnarray}
\phi_{-1}(\omega) &=&\tanh(\omega)\,, \label{eq:new.B15} \\
\phi_n^\prime(\omega) &=& \phi_{n-1}(\omega)\,, \label{eq:new.B16}\\
\phi_n(-\omega) &=& (-1)^n\phi_n(\omega)  \,, \label{eq:new.B17} \\
\phi_n(0) &=& \frac{1}{n!}\Omega^\prime_\pm(0,n)\,,\quad n=0,2,4,\dots
\label{eq:new.B18}
\end{eqnarray}
The functions $\phi_n$ for non negative $n$ follow from
eq.~(\ref{eq:new.B16}), $\phi_n(\omega)=\phi_n(0)+\int_0^\omega
dz\,\phi_{n-1}(z)$, where the integration path lies on $\C_p$. For odd
$n$, $\phi_n(0)=0$ due to parity, eq.~(\ref{eq:new.B17}). For even
$n$, $\phi_n(0)$ is well defined from eq.~(\ref{eq:new.B18}) due to
eq.~(\ref{eq:new.B8a}).

The relation between the functions $\phi_n(\omega)$ and
$\Omega^\prime_\pm(\omega,n)$ can be established by starting from
eq.~(\ref{eq:new.B9}) and recursively applying $\int_0^\omega
d\omega$. This yields
\begin{equation}
\frac{1}{n!}\Omega^\prime_\sigma(\omega,n)
=\phi_n(\omega)-\sigma P_{n+1}(\omega) \,,  \quad n=0,1,2,\dots\,, \quad\hbox{on $\C_p$}\,.
\label{eq:new.B19}
\end{equation}
Where $P_n(\omega)$ are polynomials of degree $n$ which satisfy
\begin{eqnarray}
P_0(\omega) &=& 1\,, \label{eq:new.B20} \\
P_{n+1}^\prime(\omega) &=& P_n(\omega)\,, \label{eq:new.B21}\\
P_n(-\omega) &=& (-1)^n P_n(\omega) \,, \label{eq:new.B22} \\
P_n(0) &=& \frac{(i\pi)^n}{n!}B_n(\small{\frac{1}{2}})\,,\quad n=0,1,2,\dots
\label{eq:new.B23}
\end{eqnarray}

The lowest order relations are
\begin{eqnarray}
\Omega_\sigma^\prime(\omega,0) &=& \phi_0(\omega) -\sigma\omega\,, \label{eq:new.B24}\\
\Omega^\prime_\sigma(\omega,1) &=& \phi_1(\omega)-
\sigma\left(\frac{1}{2}\omega^2-\frac{1}{6}\left(\frac{i\pi}{2}\right)^2\right)
\,, \label{eq:new.B25} \\
\frac{1}{2!}\Omega^\prime_\sigma(\omega,2) &=& \phi_2(\omega)-
\sigma\left(\frac{1}{6}\omega^3- 
\frac{1}{6}\left(\frac{i\pi}{2}\right)^2\omega\right) 
\,.  \label{eq:new.B26}
\end{eqnarray}

These relations make explicit the transformation properties of the
functions $\Omega^\prime_\sigma(\omega,n)$ under parity, $\omega\to
-\omega$. The polynomial part is responsible for an anomalous
violation of parity. Also, it follows
\begin{equation}
\Omega^\prime_\sigma(\omega,n)= (-1)^n\Omega^\prime_{-\sigma}(-\omega,n)
\end{equation}
that reflects eq.~(\ref{eq:18b}).

Next, let us discuss the periodicity properties of these functions,
which are related to the discrete gauge transformations. Since
$\tanh(\omega)$ is a periodic function with period $i\pi$, so are the
functions $\Omega_\sigma(\omega,s)$ on $\C_\sigma$. This reflects the
trivial gauge invariance of the $\zeta$-function regularization due to
the invariance of the spectrum of the Dirac operator. Let us consider
now the periodicity on $\C_p$. Since $\Omega_\sigma(\omega,s)$ takes
the same values on $\C_\sigma$ and $\C_p$ on the half-plane
$\sigma{\rm Re}(\omega)>0$, it follows that this function is periodic
on that half-plane of $\C_p$. For the other half-plane, $\sigma{\rm
Re}(\omega)<0$, take $\omega$ in the strip $-\frac{\pi}{2}<{\rm
Im}(\omega)<\frac{\pi}{2}$ (which is common to $\C_\sigma$ and
$\C_p$). Then, the value of $\Omega_\sigma(\omega+i\pi,s)$ computed by
taking $\omega+i\pi$ in $\C_p$ will be different from the value
arrived at on $\C_\sigma$ and this difference comes solely from the
term $n=0$ in eq.~(\ref{eq:new.B2}). For instance, for $\sigma=+1$,
the contribution of this term on $\C_\sigma$ would be
$(\omega+i\pi-\omega_0)_\pi^s$, whereas on $\C_p$ the contribution is
instead $(\omega+i\pi-\omega_0)_{-\pi}^s$. Noting that
$\Omega_\sigma(\omega+i\pi,s)=\Omega_\sigma(\omega,s)$ on $\C_\sigma$
and also that $\Omega_\sigma(\omega,s)$ takes the same value on
$\C_\sigma$ and $\C_p$ because of $-\frac{\pi}{2}<{\rm
Im}(\omega)<\frac{\pi}{2}$, it is easily established that, for
$\omega+i\pi\in\C_p$,
\begin{equation}
\Omega_\sigma(\omega+i\pi,s)-\Omega_\sigma(\omega,s) =(e^{-\sigma 2\pi
is}-1)(\omega+\omega_0)^s_\sigma\Theta(-\sigma m)\,.
\end{equation}
Here $\omega_0=i\pi/2$ and $m={\rm Re}(\omega)$. After analytical
extension beyond the strip $-\frac{\pi}{2}<{\rm
Im}(\omega)<\frac{\pi}{2}$, this relation holds on the whole plane
$\C_p$. From here it follows the useful relation
\begin{equation}
\Omega^\prime_\sigma(\omega+\omega_0,n)-
\Omega^\prime_\sigma(\omega-\omega_0,n)
=-\sigma 2\pi i\,\omega^n\Theta(-\sigma m) \,,\quad \omega\in\C_p,\quad
n\in{\rm Z}\,.
\end{equation}
Where $\Theta(x)$ is the step function. Using that, on $\C_p$,
$\phi_n(\omega)$ is the component of
$\Omega^\prime_\sigma(\omega,n)/n!$ which is even in $\sigma$, we also
find
\begin{equation}
\phi_n(\omega+\omega_0)-\phi_n(\omega-\omega_0)=
i\pi \,\frac{\omega^n}{n!}\varepsilon(m) \,,\quad \omega\in\C_p,\quad
n= 0,1,2,\dots\,,
\end{equation}
where $\varepsilon(x)$ stands for the sign of $x$.

Finally, let us study the zero temperature limit properties of these
functions. This corresponds to the large $\omega$ limit. Again we start
with $\omega$ in the strip $-\frac{\pi}{2}<{\rm
Im}(\omega)<\frac{\pi}{2}$, then from eq.~(\ref{eq:new.B12}) it
follows that
\begin{equation}
\Omega^\prime_\sigma(\omega,n) \sim_{\omega\to\infty}
-2\sigma\frac{\omega^{n+1}}{n+1}\Theta(-\sigma m) \,,\quad \omega\in\C_p,\quad
n= 0,1,2,\dots
\end{equation}
By analytical extension, this relation holds on $\C_p$. Moreover, for
positive $\sigma m$, the zero limit has only exponentially decreasing
corrections. Then, from eq.~(\ref{eq:new.B19}) it follows that
\begin{equation}
\phi_{n-1}(\omega) \sim_{\omega\to\infty} \varepsilon(m)
 P_n(\omega) \,,\quad \omega\in\C_p,\quad n= 0,1,2,\dots
\end{equation}
and in particular
\begin{equation}
\phi_{n-1}(\omega) =
\frac{\omega^{n}}{n!}\varepsilon(m)\left(1+
O\left(\frac{1}{\omega}\right)\right) 
 \,,\quad \omega\in\C_p,\quad
n= 0,1,2,\dots
\end{equation}


\begin{references}

\bibitem{Lykken:1991vb}
J.D.~Lykken, J.~Sonnenschein and N.~Weiss,
Int. J. Mod. Phys. {\bf A6} (1991) 1335.

\bibitem{Ginsparg:1980ef}
P.~Ginsparg,
Nucl. Phys. {\bf B170} (1980) 388;
T.~Appelquist and R.D.~Pisarski,
Phys. Rev. {\bf D23} (1981) 2305;
S.~Nadkarni,
Phys. Rev. {\bf D27} (1983) 917;
N.P.~Landsman,
Nucl. Phys. {\bf B322} (1989) 498.

\bibitem{Blau:1988kv}
S.~Blau, M.~Visser and A.~Wipf,
Nucl. Phys. {\bf B310} (1988) 163;
N.D.~Birrel and P.C.W.~Davis,
{\em Quantum fields in a curved space},
Cambridge University Press (Cambridge, England, 1982).

\bibitem{Hochberg:1998sv}
D.~Hochberg, C.~Molina-Par\'{\i}s, J.~P\'erez-Mercader and M.~Visser,
Phys. Rev. Lett. {\bf 81} (1998) 4802.

\bibitem{Deser:1982vy}
S.~Deser, R.~Jackiw and S.~Templeton,
Phys. Rev. Lett. {\bf 48} (1982) 975;
Ann. Phys. {\bf 140} (1982) 372.

\bibitem{Coleman:1985zi}
S.~Coleman and B.~Hill,
Phys. Lett. {\bf 159B} (1985) 184;
A.~Khare, R.~MacKenzie and M.B.~Paranjape,
Phys. Lett. {\bf B343} (1995) 239.

\bibitem{Wess:1971cm}
J.~Wess and B.~Zumino,
Phys. Lett. {\bf 37B} (1971) 95.

\bibitem{Witten:1983tw}
E.~Witten,
Nucl. Phys. {\bf B223} (1983) 422.

\bibitem{Redlich:1984kn}
A.N.~Redlich,
Phys. Rev. Lett. {\bf 52} (1984) 18;
Phys. Rev. {\bf D29} (1984)  2366.

\bibitem{Niemi:1983rq}
A.J.~Niemi and G.W.~Semenoff,
Phys. Rev. Lett. {\bf 51} (1983) 2077.

\bibitem{Alvarez-Gaume:1984ig}
L.~Alvarez-Gaume and E.~Witten,
Nucl. Phys. {\bf B234} (1984) 269;
R.~Jackiw,
``Topological Investigations Of Quantized Gauge Theories,"
{\it Presented at Les Houches Summer School on Theoretical Physics:
                  Relativity Grops and Topology, Les Houches, France, Jun 27
                  - Aug 4, 1983}.

\bibitem{Adler:1969av}
S.L.~Adler,
Phys. Rev. {\bf 177} (1969) 2426;
J.S.~Bell and R.~Jackiw,
Nuovo Cim. {\bf 60A} (1969) 47;
W.A.~Bardeen,
Phys. Rev. {\bf 184} (1969) 1848;
R.~Jackiw, in
S.~Treiman, R.~Jackiw, B.~Zumino and E.~Witten,
{\em Current algebra and anomalies},
World Scientific (Singapore, 1995).

\bibitem{Alvarez-Gaume:1985dr}
L.~\'Alvarez-Gaum\'e and P.~Ginsparg,
Ann. Phys. {\bf 161} (1985) 423;
Erratum-ibid.{\bf 171} (1986) 233. 

\bibitem{Alvarez-Gaume:1985nf}
L.~\'Alvarez-Gaum\'e, S.~Della Pietra and G.~Moore,
Ann. Phys. {\bf 163} (1985) 288.

\bibitem{Forte:1987em}
S.~Forte,
Nucl. Phys. {\bf B288} (1987) 252.

\bibitem{Niemi:1985ux}
A.J.~Niemi,
Nucl. Phys. {\bf B251} (1985) 155.

\bibitem{Niemi:1986vz}
A.J.~Niemi and G.W.~Semenoff,
Phys. Rept. {\bf 135} (1986) 99.

\bibitem{Sissakian:1997cp}
A.N.~Sissakian, O.Y.~Shevchenko and S.B.~Solganik,
Nucl. Phys. {\bf B518} (1998) 455.

\bibitem{Redlich:1985md}
A.N.~Redlich and L.C.~Wijewardhana,
Phys. Rev. Lett. {\bf 54} (1985) 970;
K.~Tsokos,
Phys. Lett. {\bf 157B} (1985) 413;
A.J.~Niemi and G.W.~Semenoff,
Phys. Rev. Lett. {\bf 54} (1985) 2166;
A.J.~Niemi,
Phys. Rev. Lett. {\bf 57} (1986) 1102;
A.R.~Rutherford,
Phys. Lett. {\bf 182B} (1986) 187;
J.~McCarthy and A.~Wilkins,
Phys. Rev. {\bf D58} (1998) 085007.

\bibitem{Seeley:1967ea}
R.T.~Seeley,
Proc. Symp. Pure Math. {\bf 10} (1967) 288;
J.S.~Dowker and R.~Critchley,
Phys. Rev. {\bf D13} (1976) 3224;
S.W.~Hawking,
Commun. Math. Phys. {\bf 55} (1977) 133;
E.~Elizalde, S.D.~Odintsov, A.~Romeo, A.A.~Bytsenko and S.~Zerbini,
{\em Zeta Regularization Techniques with Applications},
World Scientific (Singapore, 1994).

\bibitem{Atiyah:1975}
M.~Atiyah, V. Patodi and I.~Singer,
Proc. Camb. Phil. Soc. {\bf 77} (1975) 43;
{\bf 78} (1975) 405;
{\bf 79} (1976) 71;
T.~Eguchi, P.B.~Gilkey and A.J.~Hanson,
Phys. Rept. {\bf 66} (1980) 213.

\bibitem{GamboaSaravi:1996aq}
R.E.~Gamboa Sarav\'{\i}, G.L.~Rossini and F.A.~Schaposnik,
Int. J. Mod. Phys. {\bf A11} (1996) 2643.

\bibitem{Salcedo:1996qy}
L.L.~Salcedo and E.~Ruiz Arriola,
Ann. Phys. {\bf 250} (1996) 1.

\bibitem{Ishikawa:1987zi}
K.~Ishikawa and T.~Matsuyama,
Nucl. Phys. {\bf B280} (1987) 523;
K.S.~Babu, A.~Das and P.~Panigrahi,
Phys. Rev. {\bf D36} (1987) 3725;
E.R.~Poppitz,
Phys. Lett. {\bf B252} (1990) 417;
L.~Moriconi,
Phys. Rev. {\bf D44} (1991) 2950;
M.~Burgess,
Phys. Rev. {\bf D44} (1991) 2552.

\bibitem{Aitchison:1993md}
I.J.~Aitchison, C.D.~Fosco and J.A.~Zuk,
Phys. Rev. {\bf D48} (1993) 5895;
I.J.~Aitchison and J.A.~Zuk,
Ann. Phys. {\bf 242} (1995) 77.

\bibitem{Pisarski:1987gq}
R.D.~Pisarski,
Phys. Rev. {\bf D35} (1987) 664.

\bibitem{Bralic:1996uj}
N.~Bralic, C.D.~Fosco and F.A.~Schaposnik,
Phys. Lett. {\bf B383} (1996) 199;
D.~Cabra, E.~Fradkin, G.L.~Rossini and F.A.~Schaposnik,
Phys. Lett. {\bf B383} (1996) 434.

\bibitem{Dunne:1997yb}
G.~Dunne, K.~Lee and C.~Lu,
Phys. Rev. Lett. {\bf 78} (1997) 3434.

\bibitem{Deser:1997nv}
S.~Deser, L.~Griguolo and D.~Seminara,
Phys. Rev. Lett. {\bf 79} (1997) 1976;
Phys. Rev. {\bf D57} (1998) 7444.

\bibitem{Fosco:1997ei}
C.~Fosco, G.L.~Rossini and F.A.~Schaposnik,
Phys. Rev. Lett. {\bf 79}, 1980 (1997);
Erratum ibid.{\bf 79} (1997) 4296;
Phys. Rev. {\bf D56} (1997) 6547.

\bibitem{Fosco:1998cq}
C.D.~Fosco, G.L.~Rossini and F.A.~Schaposnik,
``Induced parity breaking term in arbitrary odd dimensions at finite
                  temperature,"
hep-th/9810199.

\bibitem{Deser:1997fu}
S.~Deser, L.~Griguolo and D.~Seminara,
Commun. Math. Phys. {\bf 197} (1997) 443;
I.J.~Aitchison and C.D.~Fosco,
Phys. Rev. {\bf D57} (1998) 1171;
R.~Jackiw and S.Y.~Pi,
Phys. Lett. {\bf B423} (1997) 364;
A.~Das and G.~Dunne,
Phys. Rev. {\bf D57} (1998) 5023;
J.~Barcelos-Neto and A.~Das,
Phys. Rev. {\bf D58} (1998) 085022;
A.~Das and A.J.~da Silva,
``Exact effective action for (1+1)-dimensional fermions in an Abelian
                  background at finite temperature,"
hep-th/9808027;
M.~Hott and G.~Metikas,
``Effective action for QED in (2+1)-dimensions at finite temperature,"
hep-ph/9812386;
C.D.~Fosco, R.E.~Gamboa Sarav\'{\i} and F.A.~Schaposnik,
``On the two-dimensional fermion determinant at finite temperature,"
hep-th/9902002.

\bibitem{Felipe:1997nx}
R.~Gonz\'alez Felipe,
Phys. Lett. {\bf B417} (1997) 114.

\bibitem{Salcedo:1998tg}
L.L.~Salcedo,
Phys. Rev. {\bf D58} (1998) 125007.

\bibitem{Pletnev:1998yu}
N.G.~Pletnev and A.T.~Banin,
``Covariant technique of derivative expansion of one loop effective action.
                  1,"
hep-th/9811031.

\bibitem{RuizArriola:1998zi}
E.~Ruiz Arriola and L.L.~Salcedo,
``Chiral and scale anomalies of nonlocal Dirac operators,"
hep-th/9811073.
To appear in Phys. Lett. {\bf B}.

\bibitem{GomezNicola:1994vq}
A.~G\'omez Nicola and R.F.~\'Alvarez-Estrada,
Int. J. Mod. Phys. {\bf A9} (1994) 1423.

\bibitem{Alvarez-Estrada:1993jm}
R.F.~\'Alvarez-Estrada, A.~Dobado and A.~G\'omez Nicola,
Phys. Lett. {\bf B319} (1993) 238.

\bibitem{Pisarski:1996ne}
R.D.~Pisarski,
Phys. Rev. Lett. {\bf 76} (1996) 3084.

\bibitem{Rothe:1986br}
K.D.~Rothe,
Nucl. Phys. {\bf B269} (1986) 269.

\bibitem{Gradshteyn:1980}
I.S.~Gradshteyn and I.M.~Ryzhik,
{\em Table of integrals, series, and products},
Academic Press, (New York, 1980).



\end{references}
\end{document}